\documentclass[fleqn,usenatbib]{mnras}
\usepackage{amsmath}
\usepackage{newtxtext,newtxmath}

\usepackage[T1]{fontenc}

\usepackage{orcidlink}
\usepackage{hyperref} 
\DeclareRobustCommand{\VAN}[3]{#2}
\let\VANthebibliography\thebibliography
\def\thebibliography{\DeclareRobustCommand{\VAN}[3]{##3}\VANthebibliography}

\usepackage{graphicx}	
\usepackage{amsmath}	

\usepackage{xcolor}
\usepackage[normalem]{ulem}

\newcommand{\HI}{\ion{H}{i}}

\title[21CMA Direction-Dependent Source Catalogue]{A Point-source Catalogue from Direction-dependent Calibrated 21CMA Images}
\author[Qizhe Tang et al.]{
Qizhe Tang$^{1,2}$\orcidlink{0009-0006-8400-3699},
Jingying Wang$^{1}$\thanks{E-mail: jywang@shao.ac.cn}\orcidlink{0000-0002-5598-2668}, 
Junhua Gu$^{3}$\orcidlink{0000-0001-9765-6521},
Qian Zheng$^{1}$\orcidlink{0000-0001-5345-6050},
Quan Guo$^{1}$\orcidlink{0000-0003-2858-5090},
Tingyan Shen$^{1}$\orcidlink{0000-0001-8688-0675},
\and
Bixuan Zhao$^{1}$\orcidlink{0000-0002-1007-1991},
Qiuyi Lai$^{1,2}$\orcidlink{0009-0003-0945-0358},
Yan Huang$^{3}$\orcidlink{0000-0002-5008-173X},
Kuanjun Li$^{3}$\orcidlink{0009-0006-3262-0775},
Yang Feng$^{3}$\orcidlink{0009-0008-7696-5374} \\
\\ $^{1}$State Key Laboratory of Radio Astronomy and Technology, Shanghai Astronomical Observatory, CAS, 80 Nandan Road, Shanghai 200030, China\\
$^{2}$School of Astronomy and Space Science, University of Chinese Academy of Sciences, Beijing 100049, China\\
$^{3}$State Key Laboratory of Radio Astronomy and Technology, National Astronomical Observatories, CAS, 20A Datun Road, Beijing 100101, China
}

\date{Accepted XXX. Received YYY; in original form ZZZ}

\pubyear{\the\year{}}

\begin{document}
\label{firstpage}
\pagerange{\pageref{firstpage}--\pageref{lastpage}}
\maketitle

\begin{abstract}
We present a deep low-frequency radio point-source catalogue 
centred on the North Celestial Pole (NCP) and extending $5^\circ$ in radius, 
derived from a single 24-h observation with the 21 CentiMeter Array (21CMA). 
Covering 75--175\,MHz, the catalogue characterises extragalactic foreground point sources for Epoch of Reionization (EoR) studies. 
Previous 21CMA imaging has been limited by reduced dynamic range caused by direction-dependent effects, particularly the severe grating lobes of ultra-bright sources such as 3C\,61.1 and ionospheric phase distortions. 
Here we apply \texttt{21CMAcali}, a direction-dependent calibration and imaging pipeline, 
to mitigate these effects. 
The angular resolution ranges from 2.3\,arcmin at 100\,MHz to 1.4\,arcmin at 168.75\,MHz. 
The resulting maps show reduced artefacts and background noise, enabling more reliable recovery of faint sources. 
We construct a catalogue of 1246 radio point sources. 
Cross-matching with LoTSS-DR3, WENSS, and NVSS enables broad-band spectral fitting, while an independent in-band analysis yields a median spectral index of $\alpha=-0.825$ (where $S \propto \nu^{\alpha}$). 
The Euclidean-normalised differential source counts at 150\,MHz, 
derived from a dedicated 50-MHz-wide multi-frequency synthesis (MFS) image, 
are broadly consistent with existing measurements and semi-empirical models at the bright end, whereas 
the deficit at lower flux densities ($S \lesssim 0.04\,\mathrm{Jy}$) 
mainly reflects limited sensitivity and incompleteness. 
This catalogue improves foreground point-source characterisation in the 21CMA NCP field, demonstrates the value of direction-dependent calibration, and provides a reference for deeper catalogues based on multi-day observations.
\end{abstract}




\section{Introduction}
\label{sec:introduction}
The Epoch of Reionization (EoR) marks the period in cosmic history, during which ultraviolet and X-ray radiation from the first stars and galaxies ionized the neutral hydrogen (\HI) in the intergalactic medium \citep{2013ASSL..396...45Z}. The redshifted 21-cm signal from the neutral hydrogen, observed at low radio frequencies ($\sim 50\text{--}200\,\mathrm{MHz}$) over the corresponding redshift range of $z \approx 6\text{--}27$, provides a promising probe of this era. It can be used to trace the timing and duration of reionization, constrain the properties of the first ionizing sources, and probe the large-scale structure of the early Universe \citep[e.g.,][]{2006PhR...433..181F, 2012RPPh...75h6901P, CD-SKA, EoR-review-1, 21-cm_revolution, 2020PASP..132f2001L, EoR-review-2, 21-cm_line}.

Detecting the redshifted 21-cm signal from the EoR is a major goal of current low-frequency radio experiments. Instruments such as the Giant Metrewave Radio Telescope (GMRT; \citealp{GMRT-EoR}), the Low Frequency Array (LOFAR; \citealp{LOFAR}), the Murchison Widefield Array (MWA; \citealp{MWA}), the Precision Array for Probing the Epoch of Reionization (PAPER; \citealp{PAPER}), and the Hydrogen Epoch of Reionization Array (HERA; \citealp{HERA}) have made substantial progress in this direction and laid the groundwork for the Square Kilometre Array (SKA; \citealp{SKA}). However, the expected EoR signal, with a typical brightness temperature of only $\sim 10\text{--}30\,\mathrm{mK}$, is overwhelmed by foreground emission that is four to five orders of magnitude brighter \citep{2008MNRAS.388..247D,2010ApJ...723..620W,2013ApJ...763...90W, 2020PASP..132f2001L}. These foregrounds include diffuse Galactic synchrotron emission and a large population of extragalactic compact radio sources, whose coupling to the instrumental response can generate systematics that mimic or obscure the 21-cm signal \citep{2010ARA&A..48..127M}. Accurate characterisation of the foreground point-source population is therefore a key requirement for EoR studies.

The 21 CentiMeter Array (21CMA; \citealp{21CMA}) is a dedicated low-frequency interferometric array for EoR studies and has carried out long-term observations of the North Celestial Pole (NCP) region. Over years of operation, it has accumulated an extensive archival dataset covering 50--200\,MHz. Based on these observations, previous studies have produced radio-source catalogues in the NCP field. Using approximately 12 h of data obtained on 2013 April 13, \citet{Zheng2016} selected east-west baselines with lengths of 100--1500\,m within the 75--175\,MHz band and released the first 21CMA source catalogue of this region, containing 624 radio sources. Subsequently, \citet{Zhao2022} combined data from 2013 May 10, 13, and 22 and, using all east-west baselines shorter than 2780\,m over the 75--200\,MHz band, increased the number of detected point sources to 832.

However, these earlier studies were primarily based on direction-independent (DI) calibration. As a result, high-dynamic-range imaging remained limited by the grating lobes of ultra-bright radio sources in the field, particularly 3C\,61.1, as well as by residual ionospheric phase errors. In this work, we apply \texttt{21CMAcali}\footnote{\url{https://github.com/astroJyWang/21CMAcali/}}, a direction-dependent calibration and wide-field imaging pipeline, to reprocess a continuous 24-h dataset obtained on 2013 March 25. By partitioning the wide field into multiple directions (facets) and correcting for spatially varying ionospheric and primary-beam effects, the direction-dependent processing improves the fidelity and dynamic range of the restored images and enables more reliable source recovery in the NCP field. Based on these images, we construct a new point-source catalogue covering the central $5^\circ$ region around the NCP. Compared with previous 21CMA catalogues, the present catalogue improves the astrometric accuracy by more than an order of magnitude and provides a more complete recovery of faint sources affected by dynamic-range limitations in earlier imaging.

This paper is organised as follows. In Section \ref{sec:data}, we describe the 21CMA observations and the data reduction pipeline, including the electromagnetic beam simulations and the direction-dependent calibration strategy. Section \ref{sec:point source finding} presents the point-source extraction procedure based on local background estimation and morphological filtering. Section \ref{sec:rescale} describes the catalogue construction and refinement process, including astrometric and flux density scale corrections, and in-band spectral fitting. Section \ref{sec:catalogue} presents the format of the final catalogue. Section \ref{sec:validation} assesses the astrometric and photometric performance, statistical reliability, and comparison with previous 21CMA results. Section \ref{sec:discussion} discusses the spectral-index distribution, intrinsic noise properties, detection completeness, and Euclidean-normalised differential source counts. Finally, Section \ref{sec:conclusion} summarises our conclusions and briefly discusses future extensions to deeper catalogues based on the combination of multiple days of 21CMA observations.

\section{Observations and Data Reduction}\label{sec:data}

\subsection{The 21CMA}
\label{sec:21CMA}

The 21CMA, located in the Ulastai valley of the Tianshan Mountains in Xinjiang, western China ($86^\circ 41'$E, $42^\circ 56'$N; \citealp{21CMA}), is a low-frequency radio interferometric array dedicated to studies of the redshifted 21-cm neutral hydrogen signal from the EoR. The array consists of two perpendicular arms: an East-West (EW) arm extending approximately 6.1\,km and a North-South (NS) arm extending about 4.0\,km. The entire array comprises 81 stations, with 41 aligned along the EW arm and 40 along the NS arm. In the observing mode used for the present analysis, visibilities are computed from signals collected by 40 stations of the EW arm; the most distant 41st station was not used.

The 21CMA operates within a frequency range of 50--200\,MHz, corresponding to the redshift range $z \approx 6\text{--}27$ for the redshifted 21-cm signal. All antenna elements of the 21CMA are fixed on the ground and point towards the NCP, allowing long-integration observations of the same field as the Earth rotates. In this work, we focus on the region within about $5^\circ$ of the NCP, which roughly corresponds to the field of view set by the primary beam of a single station. 

\begin{figure}
    \centering
    \includegraphics[width=\columnwidth]{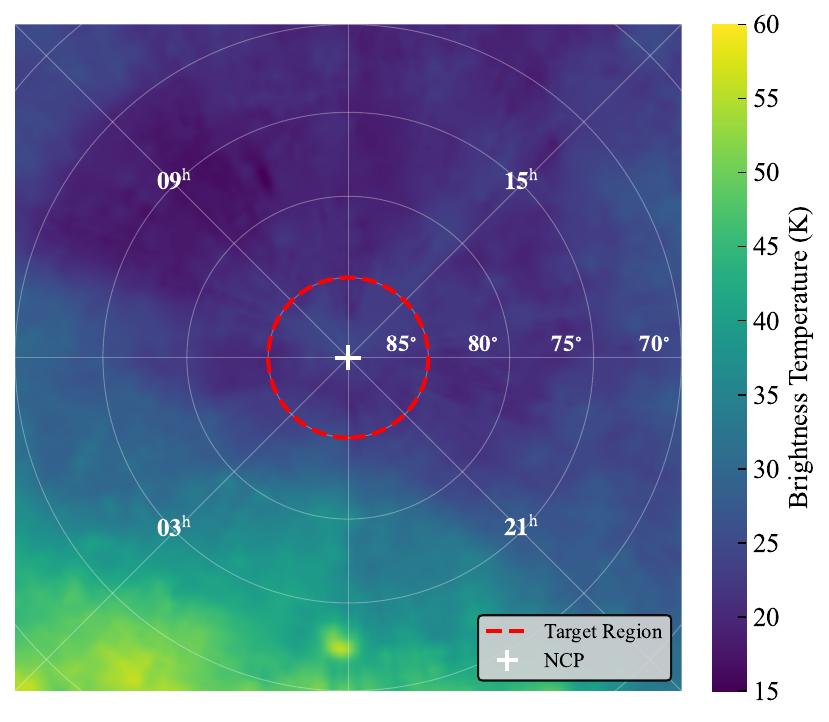}
    \caption{The 21CMA target region projected on to the Haslam 408\,MHz sky map \citep{Haslam-1, Haslam-2}. The image covers the region around the NCP for $\mathrm{Dec} > 70^\circ$. The colour bar shows the brightness temperature in K. The white cross marks the NCP, and the red dashed circle indicates the core target region ($\mathrm{Dec} > 85^\circ $)  in this work.}
    \label{fig:haslam_map}
\end{figure}

To illustrate the large-scale radio background of the observed field, Figure \ref{fig:haslam_map} shows the 21CMA target region projected on to the Haslam 408\,MHz all-sky map \citep{Haslam-1, Haslam-2}. The selected NCP field lies well away from the bright synchrotron emission of the Galactic plane, making it a relatively favourable region for EoR studies.

\subsection{Primary Beam Modelling}
\label{sec:primary_beam}

The 21CMA operates on the principle of aperture synthesis, with the stations forming the interferometric pairs used to synthesise a large effective aperture. Each station consists of a dense hexagonal cluster of 127 single-polarised log-periodic antennas, giving a total of 10,287 antenna elements across the array.

To characterise the direction-dependent response of the array, we do not adopt the Gaussian beam models used in previous studies \citep[e.g.,][]{Zheng2016}. Although a circularly symmetric Gaussian model can provide a reasonable approximation to the inner main lobe, its in situ calibration is constrained by the limited number and spatial distribution of available in-field calibrators, and it does not capture the angular asymmetry around the beam center and complex sidelobe structure of the beam. In addition, the relatively low frequency resolution ($\sim 1.56$\,MHz) of previous models tends to smooth over the fine spectral features associated with antenna resonances.

Instead, we adopt a numerical electromagnetic simulation of the beam. Using the \textsc{NEC2++} code \citep{Molteno2014NEC2}, based on the Method of Moments, we construct a three-dimensional physical model of a single log-periodic antenna, including the transmission lines and ground plane. The full station beam is then synthesised by coherently combining the 127 antenna signals, each compensated with a geometric delay to ensure the beam points towards the NCP. The resulting beam model is generated at a spectral resolution of $\sim 97.6$\,kHz and evaluated on a full-sky HEALPix grid to avoid additional projection effects.

\begin{figure}
	\centering
	\includegraphics[width=\columnwidth]{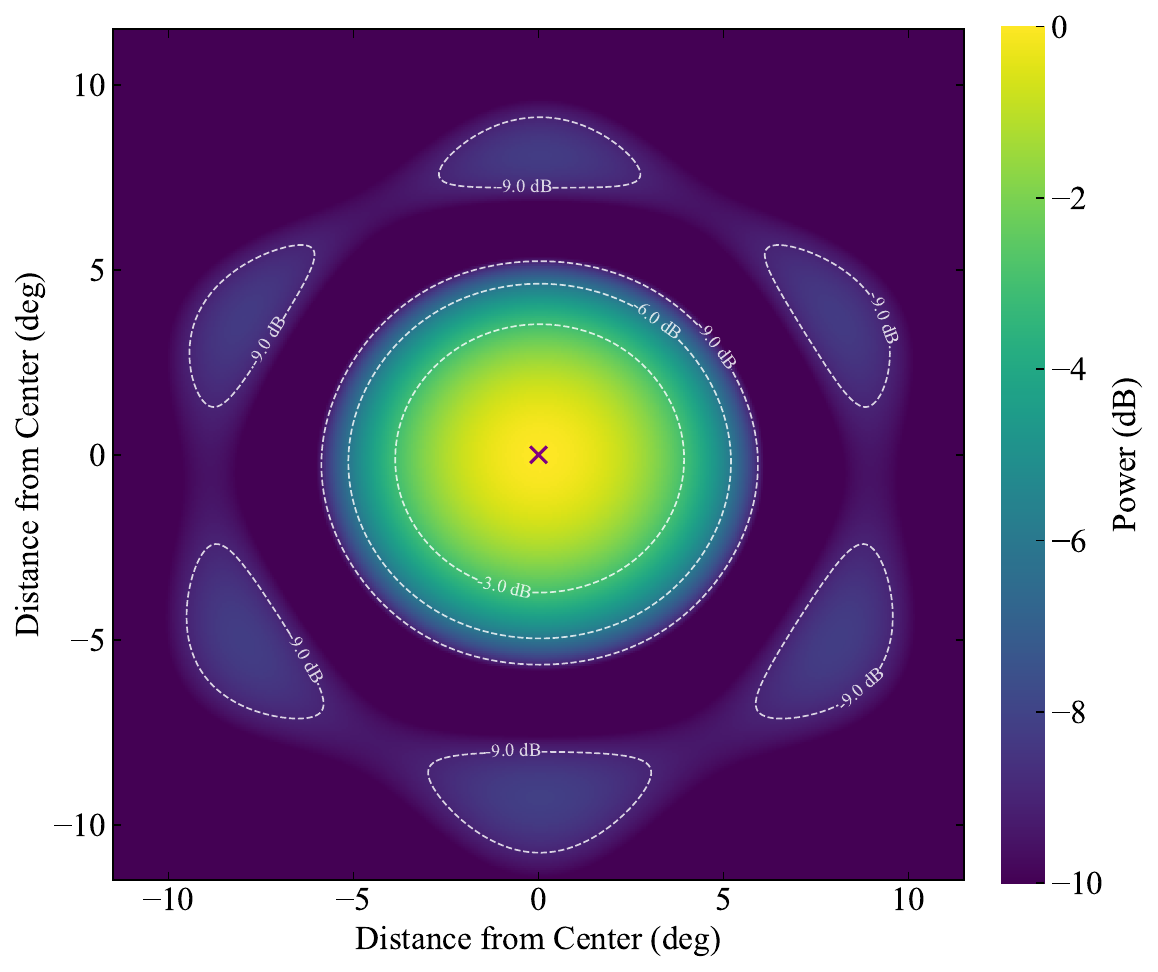}
	\caption{Simulated two-dimensional normalised primary beam power pattern of a single 21CMA station at 150\,MHz. The model is generated using the \textsc{NEC2++} code \citep{Molteno2014NEC2}, taking into account the geometry of the 127 log-periodic antennas arranged in a hexagonal cluster. The small cross marks the beam centre, corresponding to the NCP pointing direction.}
	\label{fig:beam_150MHz}
\end{figure}

Figure \ref{fig:beam_150MHz} shows the simulated two-dimensional normalised primary beam power pattern of a single station at 150\,MHz. To characterise the frequency dependence of the main lobe over the 75--175\,MHz band, we uniformly sampled 32 frequency slices from the instantaneous \textsc{NEC2++} beam models. For each slice, we calculated the primary beam averaged over angles about the beam center, which approximates the time-integrated response of a 24-h full-synthesis observation centred on the NCP, and measured its full width at half-maximum (FWHM).

\begin{figure}
    \centering
    \includegraphics[width=\columnwidth]
    {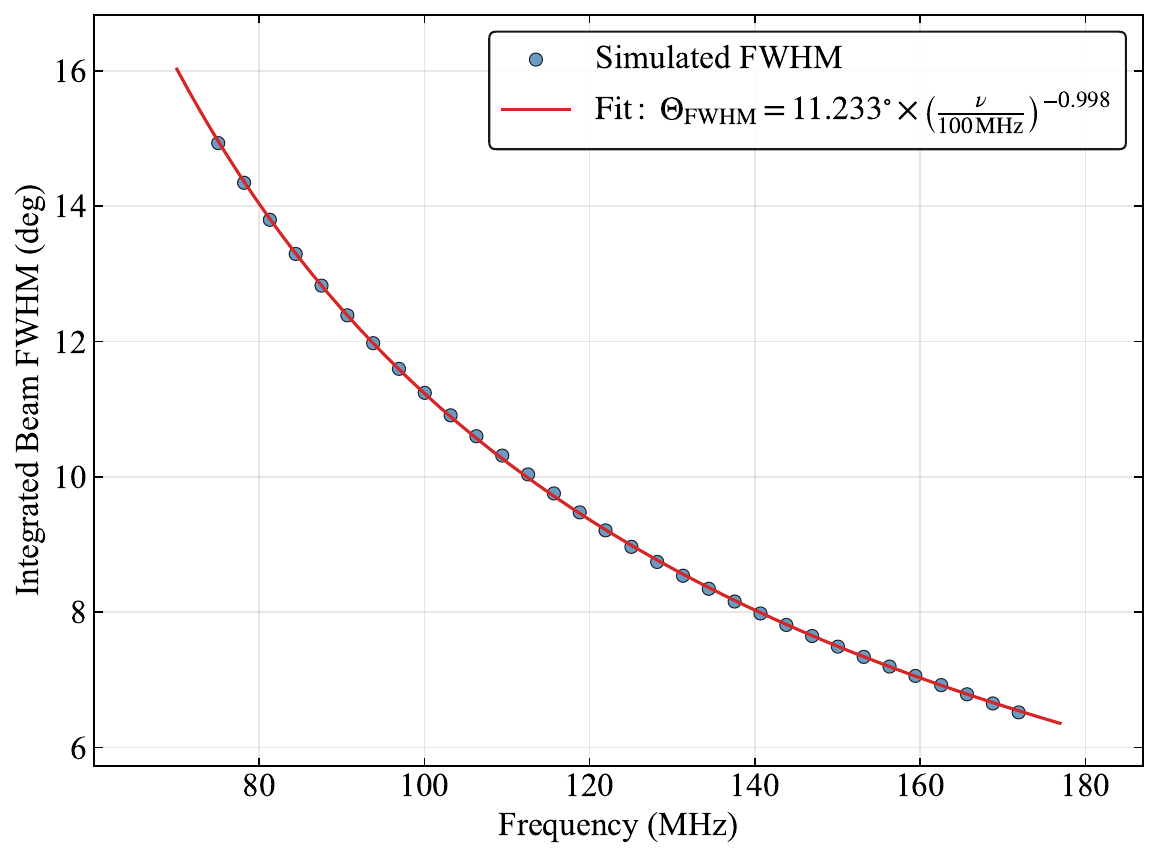}
    \caption{FWHM of the integrated primary beam of a single 21CMA
    station as a function of frequency. The blue circles show the FWHM
    values measured from 32 uniformly sampled frequency slices between
    75 and 175\,MHz based on the \textsc{NEC2++} simulations. The red
    solid line shows the best-fitting power-law model obtained with a
    least-squares fit.}
    \label{fig:FWHM}
\end{figure}

As shown in Figure \ref{fig:FWHM}, the FWHM of the integrated beam decreases with frequency. At 150\,MHz, the strongest near sidelobe in the azimuthally averaged 24-h integrated beam profile peaks at approximately $-8.1$\,dB relative to the main-lobe maximum, corresponding to 15.5 per cent of the main-lobe peak power.
We fit the extracted FWHM measurements with a power-law model, $\Theta_{\mathrm{FWHM}} = A (\nu / 100\,\mathrm{MHz})^B$, using a least-squares method. The best-fitting parameters are $A = 11.233^\circ$ at 100\,MHz and $B = -0.998$. The fitted index is also consistent with the expected scaling for a diffraction-limited aperture, $\theta \propto \lambda/D \propto \nu^{-1}$, supporting the physical consistency of the numerical beam model. This frequency-dependent beam model captures the variation of the main lobe across the band and provides the primary beam model required for the direction-dependent calibration and wide-field imaging (Section \ref{sec:dd_imaging}).

\subsection{Observation Strategy and Data Selection}
\label{sec:obs_strategy}

The 21CMA correlator produces visibility data with a frequency resolution of approximately $24.4\,\mathrm{kHz}$ and a time resolution of $3\,\mathrm{s}$. In this work, we use a continuous 24-h observation obtained on 2013 March 25. As the primary goal of this study is the extraction and analysis of compact point sources, we exclude projected baselines shorter than $100\,\mathrm{m}$, which are more strongly affected by diffuse Galactic emission and residual radio frequency interference (RFI), while retaining all longer East--West baselines up to the maximum available length of approximately $2.7\,\mathrm{km}$. 
In addition, to mitigate unstable gain variations associated with the roll-off of the analogue bandpass filters, we discard the band-edge
data at $50$--$75\,\mathrm{MHz}$ and $175$--$200\,\mathrm{MHz}$, retaining a usable bandwidth of $75$--$175\,\mathrm{MHz}$.

To improve the calibration and imaging quality, we divide the usable 75--175\,MHz bandwidth into eight 12.5\,MHz sub-bands. This strategy balances the frequency dependence of the instrumental response, the signal-to-noise ratio available for calibration, and the practical complexity of direction-dependent processing. The lower-frequency range, 75--125\,MHz, is more strongly affected by terrestrial RFI, including FM broadcasting, analogue television signals, and computer-generated noise \citep{Huang2016}. To increase the signal-to-noise ratio and compensate for the lower angular resolution in this regime, we first perform direction-independent calibration on the four lower sub-bands (75--87.5, 87.5--100, 100--112.5, and 112.5--125\,MHz), and then combine them into a single 50\,MHz block for subsequent direction-dependent calibration. By contrast, the higher-frequency range, 125--175\,MHz, has a comparatively cleaner radio environment and is processed throughout as four individual 12.5\,MHz sub-bands (125--137.5, 137.5--150, 150--162.5, and 162.5--175\,MHz). This subdivision facilitates reliable source extraction and in-band spectral fitting. The detailed calibration and imaging parameters adopted after data selection are summarised in Table \ref{tab:observing_parameters}.

\begin{table}
	\centering
	\caption{Summary of the selected data and calibration/imaging parameters used in this work.}
	\label{tab:observing_parameters}
	\begin{tabular}{lr} 
		\hline
		\textbf{Parameter} & \textbf{Value} \\
		\hline
		Observation Date & 2013 March 25 \\
		Total Integration Time & 24\,h \\
		Frequency Coverage & 75--175\,MHz \\
		\quad \textit{Low-band processing} & 75--125\,MHz (single 50\,MHz block) \\
		\quad \textit{High-band processing} & 125--175\,MHz (4 $\times$ 12.5\,MHz sub-bands) \\
		Frequency Resolution & $\sim$24.4\,kHz \\
		Time Resolution & $\sim$3\,s \\
		Baseline Range (EW only) & 100\,m--2.7\,km \\
		Instrumental Polarisation & Linear (XX) \\
		Sky Model & 137 point sources + 3C\,61.1 (2 components) \\
		Primary Beam Model & Frequency-dependent \textsc{NEC2++} simulation \\
		\hline
	\end{tabular}
\end{table}

\subsection{Sky Model Construction and Pre-processing}
\label{sec:preprocessing}

An accurate apparent-flux sky model is essential for the initial calibration. We first queried the NASA/IPAC Extragalactic Database (NED)\footnote{\url{https://ned.ipac.caltech.edu/}} for sources in the NCP region with $\mathrm{Dec} > 85^\circ$ and available radio flux-density measurements. We included objects classified as galaxies, QSOs, and radio sources, while excluding group- and cluster-like entries (e.g. galaxy pairs, triples, groups, clusters, and QSO groups). To improve the reliability of the low-frequency sample, we further restricted the selection to sources associated with the following radio catalogues: the Very Large Array Low-frequency Sky Survey Redux (VLSSr; \citealp{VLSSr}) at 73.8\,MHz, the Tata Institute of Fundamental Research Giant Metrewave Radio Telescope Sky Survey Alternative Data Release 1 (TGSS-ADR1; \citealp{TGSS_ADR1}) at 150\,MHz, the revised Eighth Cambridge survey (8C; \citealp{8C}) at 38\,MHz, the Seventh Cambridge survey (7C; \citealp{7C}) at 151\,MHz, the Sixth Cambridge survey (6C; \citealp{6C}) at 151\,MHz, and the Third Cambridge catalogue (3C; \citealp{3C}) at 159\,MHz.
This initial selection yielded 642 candidate sources. We then fitted a power-law spectrum to the available flux-density measurements below 500\,MHz for each source, iteratively rejecting outlying data points lying beyond $3\sigma$ from the fitted model and requiring at least two measurements to remain after clipping. After this filtering, 579 sources were retained for the construction of the calibration sky model. The resulting calibration catalogue records the fitted spectral index and the reference intrinsic flux density for each source. The source positions in the calibration catalogue were taken from the J2000 coordinates of the corresponding NRAO VLA Sky Survey (NVSS; \citealp{NVSS}) counterparts identified through NED.

Among them, the prominent radio galaxy 3C\,61.1 requires special treatment because of its extended and complex morphology. Based on the 150-MHz total flux density of $33.23\pm7.82$\,Jy and the spectral index $\alpha=-0.80\pm0.05$ (where $S \propto \nu^{\alpha}$) reported by \citet{Zheng2016}, the total flux density of 3C~61.1 is estimated to decrease from approximately $58$\,Jy at 75\,MHz to $29$\,Jy at 175\,MHz. The 3CRR Atlas \citep{3CRR-Atlas}, based on the 3CRR sample of \citet{3CRR}, gives a largest angular size of $186$\,arcsec, corresponding to approximately $3.1$\,arcmin.
A more complete representation would ideally use orthogonal basis functions such as \textsc{Shapelets} \citep{shapelet-1,shapelet-2}, as adopted for 3C\,61.1 in the LOFAR EoR analysis of the NCP field \citep{3C61}. However, the \textsc{CASA}/\textsc{killMS} workflow adopted in the
\texttt{21CMAcali} pipeline does not support direct
\textsc{Shapelet} model inputs. We therefore adopt a simplified representation in which the two dominant lobes of 3C\,61.1 are treated as two independent point-source components.

For the initial wide-field DI calibration, the discrete source model injected in the \textsc{CASA} workflow does not explicitly account for the position-dependent primary-beam attenuation of individual sources. We therefore construct an apparent sky model for this step. This also helps to avoid an inaccurate amplitude balance between sources near the beam centre and those farther from the pointing centre, which would otherwise degrade the overall calibration quality. Similar apparent-sky treatments have also been adopted in other wide-field low-frequency analyses \citep[e.g.,][]{2024A&A...681A..62M}. To do so, we calculate a 24-h azimuthally averaged primary beam from the instantaneous \textsc{NEC2++} beam model to represent the time-integrated response. For each source, the beam response is evaluated at its angular offset from the NCP, and the intrinsic flux density is multiplied by the beam response at each sub-band centre to obtain the apparent flux density in that sub-band. We then fit the apparent spectrum of each source and record the resulting apparent spectral index and reference apparent flux density for use in the initial calibration.

To construct a reduced apparent-flux sky model for the initial DI calibration, we applied a simple spatial filtering scheme \citep{spatial-gredding_1, spatial-gridding_2}.  The purpose of this filtering was to limit the number of model components while retaining bright sources across the target field, thereby providing a reduced sky model for the initial DI calibration. The facets and calibration directions used for the subsequent direction-dependent calibration were defined separately at a later stage, as described below. Specifically, we divided the target region into 30 sectors using $1^\circ$ bins in Declination and $60^\circ$ bins in Right Ascension. The R.~A.--Dec. bins subtend unequal sky areas near the NCP. They were used only to select sources for the initial DI sky model. Within each sector, the candidate sources were ranked by apparent flux density at 150\,MHz. If a sector contained five or fewer sources, all were retained; otherwise, only the five brightest sources were selected. This spatial filtering yielded a calibration sky catalogue containing 137 point sources together with the two components of 3C\,61.1, i.e. 139 model components in total.

The data pre-processing is organised into two successive calibration stages, corresponding to the Level 1 and Level 2 modules of \texttt{21CMAcali}. In the first stage, we flag the raw visibilities with \textsc{AOFlagger} \citep{AOFlagger} using a generic strategy to remove strong time--frequency localised interference, and then perform an initial DI calibration in \textsc{CASA} \citep{CASA} using the calibrator catalogue described above. Model visibilities are predicted with the \texttt{ft} task, and a first set of complex gain solutions is derived with \texttt{bandpass} using a solution interval of 10 min. These solutions are applied with \texttt{applycal}. In the second stage, a further pass of \textsc{AOFlagger} is carried out on the calibrated data column (\texttt{CORRECTED\_DATA}) using an MWA-optimised strategy \citep{mwa-rfi,lua}, which is more sensitive to faint residual RFI. A preliminary deconvolved image is then generated from the flagged, DI-corrected data, and the CLEAN component model derived from this image is used as the updated sky model for self-calibration. A second \texttt{bandpass} run is subsequently performed using the same 10-min solution interval to refine the DI calibration before direction-dependent processing.

\subsection{Direction-dependent Calibration and Wide-field Imaging}
\label{sec:dd_imaging}

\begin{figure*}
    \centering
    \includegraphics[width=2\columnwidth]{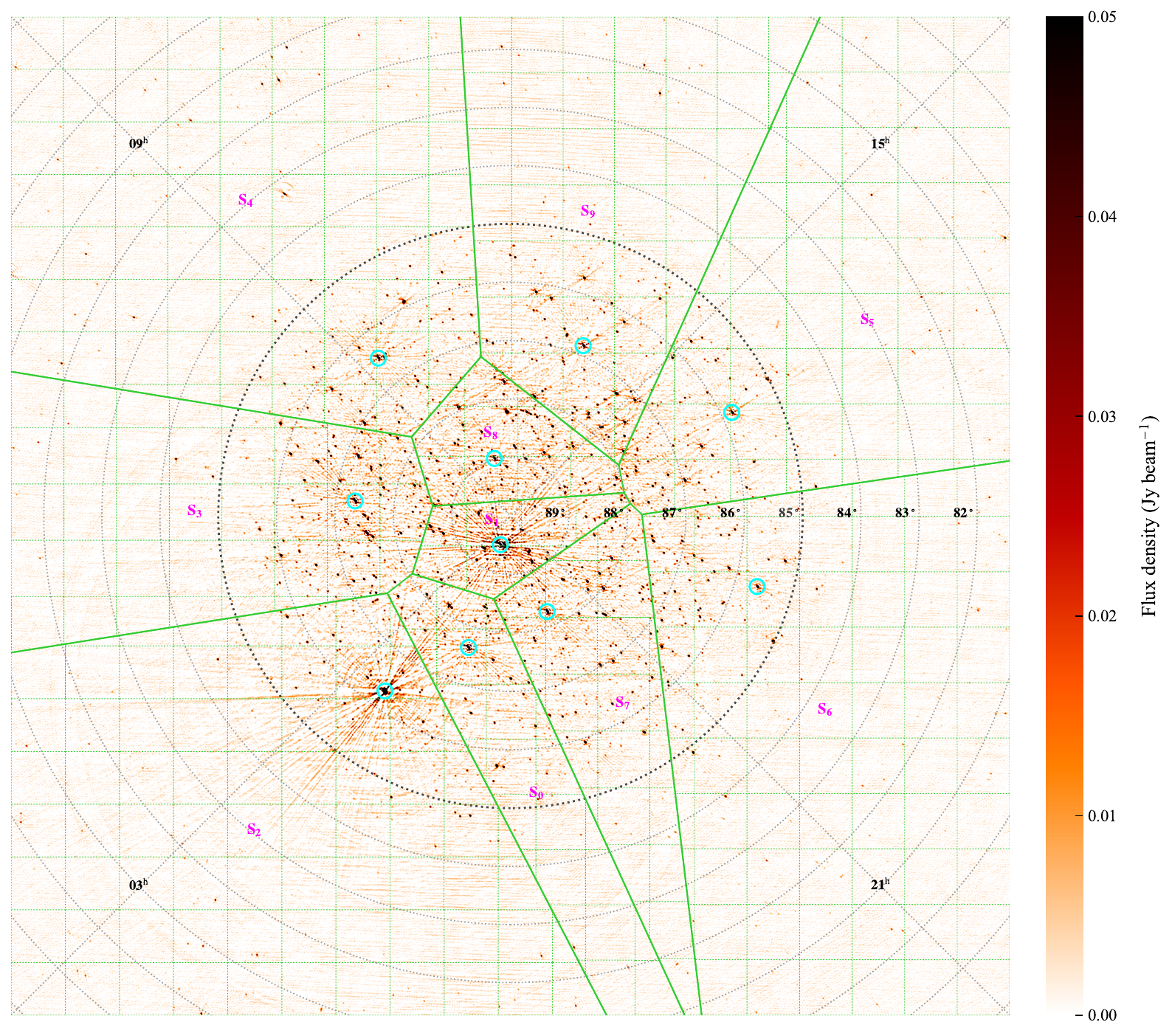}
    \caption{Directional partition used for the DD calibration of the 150--162.5\,MHz frequency product. The background shows the DI-calibrated apparent-flux image. The cyan circles mark the ten bright anchor sources within $5^\circ$ of the NCP, and the magenta labels S0--S9 identify the corresponding principal calibration facets. The thick green lines show the boundaries of these principal facets, while the finer green lines indicate the internal subdivisions used by \textsc{DDFacet} for wide-field imaging and deconvolution. The grey dashed lines show the R.~A.--Dec. coordinate grid, and the dark grey circular boundary marks $\mathrm{Dec}=85^\circ$.}
    \label{fig:dd_partition}
\end{figure*}

The direction-dependent calibration (DDC) and imaging stage (Level 3 of \texttt{21CMAcali}) is performed with the \textsc{DDFacet}/\textsc{killMS} framework \citep{2015MNRAS.449.2668S,DDFacet}. Starting from the DI-corrected visibilities in the \texttt{CORRECTED\_DATA} column, we first generate a reference dirty image with \textsc{DDFacet}. To initialise the directional calibration geometry, we select 10 bright anchor sources approximately uniformly distributed across the field and use them to define a coarse initial partition. This partition is then refined into a clustered directional model for the subsequent direction-dependent (DD) deconvolution and gain solving. The ten anchor sources define ten principal calibration facets. Finer internal subdivisions, with characteristic sizes of $\sim0.1^\circ$--$1^\circ$, are used by \textsc{DDFacet} for wide-field imaging and deconvolution but do not correspond to additional independently solved gain directions. The same ten anchor sources and facet layout are adopted for all five frequency products. All ten anchor sources lie within $5^\circ$ of the NCP; the tessellation is not truncated at the $\mathrm{Dec}=85^\circ$ catalogue boundary, so some facet polygons extend beyond this region. Figure~\ref{fig:dd_partition} shows the directional partition for the 150--162.5\,MHz frequency product.

A frequency-dependent primary-beam model derived from the \textsc{NEC2++} simulations is supplied consistently to both \textsc{DDFacet} and \textsc{killMS}. The beam response is therefore explicitly incorporated into both the deconvolution and the DD gain solving. This prevents the solver from erroneously attributing the spatial attenuation of the primary beam to instrumental or ionospheric gain variations, a process that would otherwise implicitly absorb the beam effects into the derived Jones matrices. Using the clustered directional model, we first perform a clustered deconvolution in \textsc{DDFacet} with the \texttt{SSD2} algorithm and a third-order spectral polynomial. An external mask is then generated from the restored image using a high threshold of $18\sigma$, and a deeper deconvolution pass is carried out with this mask to obtain a more reliable sky model for calibration.

Direction-dependent complex gains are subsequently solved with \textsc{killMS} using the KAFCA solver in scalar mode, with a solution time interval of 5 min and frequency interval of 1.5625 MHz. These intervals were chosen empirically to balance the need to track temporal and spectral gain variations against maintaining sufficient signal-to-noise ratio in each calibration direction. The clustered directional model thus defines multiple calibration regions, each associated with an independent set of DD gain solutions. These solutions are then applied jointly in a final \textsc{DDFacet} imaging run on the full data set. For the final catalogue images, we adopt an image size of $4096 \times 4096$ pixels with a cell size of $15$\,arcsec. Briggs weighting \citep{1995AAS...18711202B} with a robustness parameter of $-1.5$ is used. For the final imaging, we use all available East--West baselines with projected lengths from $100\,\mathrm{m}$ to the maximum available length of approximately $2.7\,\mathrm{km}$. The final restored images and residual maps from this stage are then used for subsequent source extraction and catalogue construction.

\begin{figure*}
    \centering
    \includegraphics[width=2\columnwidth]{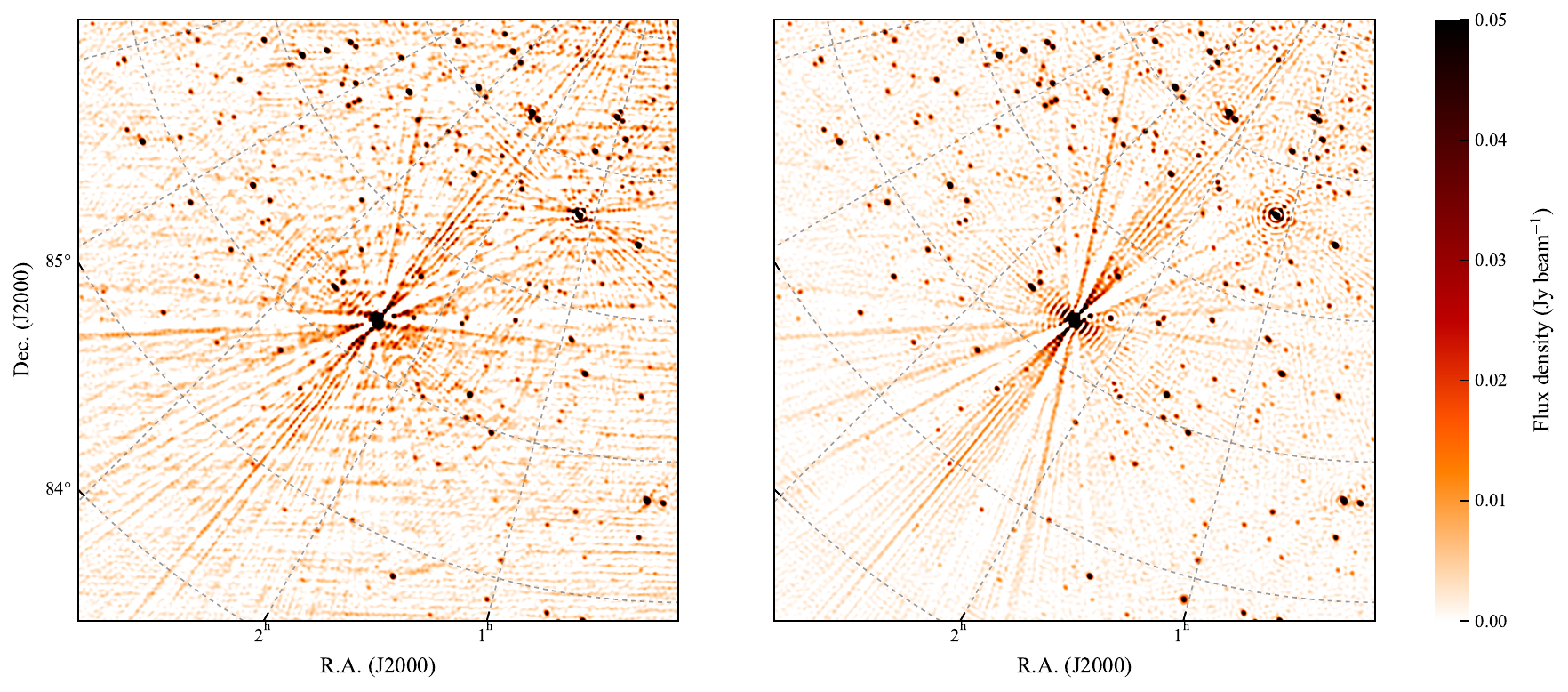}
    \caption{Comparison of the DI- and DD-calibrated apparent-flux images around 3C\,61.1 in the 150--162.5\,MHz frequency product. The left panel shows the image after DI calibration, and the right panel shows the image after DD calibration. Both images use the same visibility selection, weighting, image dimensions, pixel scale, imaging region, and displayed flux-density range. Neither image is primary-beam corrected.}
    \label{fig:di_dd_3c611}
\end{figure*}

To assess the improvement produced by the DD calibration, we generated matched DI- and DD-calibrated apparent-flux images before primary-beam correction using the same visibility selection and imaging parameters. Figure~\ref{fig:di_dd_3c611} presents a representative comparison around 3C\,61.1 in the 150--162.5\,MHz frequency product. The DD-calibrated image shows weaker radial, sidelobe-like, and other structured artefacts around the source. The corresponding dynamic ranges, measured using the same definition and declination range, are listed in Table~\ref{tab:di_dd_dynamic_range}. DD calibration improves the dynamic range in all five frequency products, with improvement factors ranging from 1.19 to 2.03.

\begin{table}
    \centering
    \caption{Dynamic ranges of the matched apparent-flux images before and after DD calibration. The peak brightness and residual rms are measured over $85^\circ\leq\mathrm{Dec}\leq90^\circ$.}
    \label{tab:di_dd_dynamic_range}
    \begin{tabular}{cccc}
        \hline
        Central frequency & DI dynamic range & DD dynamic range & DD/DI \\
        (MHz) & & & \\
        \hline
        100.00 & 1488.3 & 3021.9 & 2.03 \\
        131.25 & 1596.2 & 2378.1 & 1.49 \\
        143.75 & 2260.8 & 2755.4 & 1.22 \\
        156.25 & 2209.4 & 3229.1 & 1.46 \\
        168.75 & 4445.0 & 5302.3 & 1.19 \\
        \hline
    \end{tabular}
\end{table}

\begin{figure*}
    \centering
    \includegraphics[width=2\columnwidth]{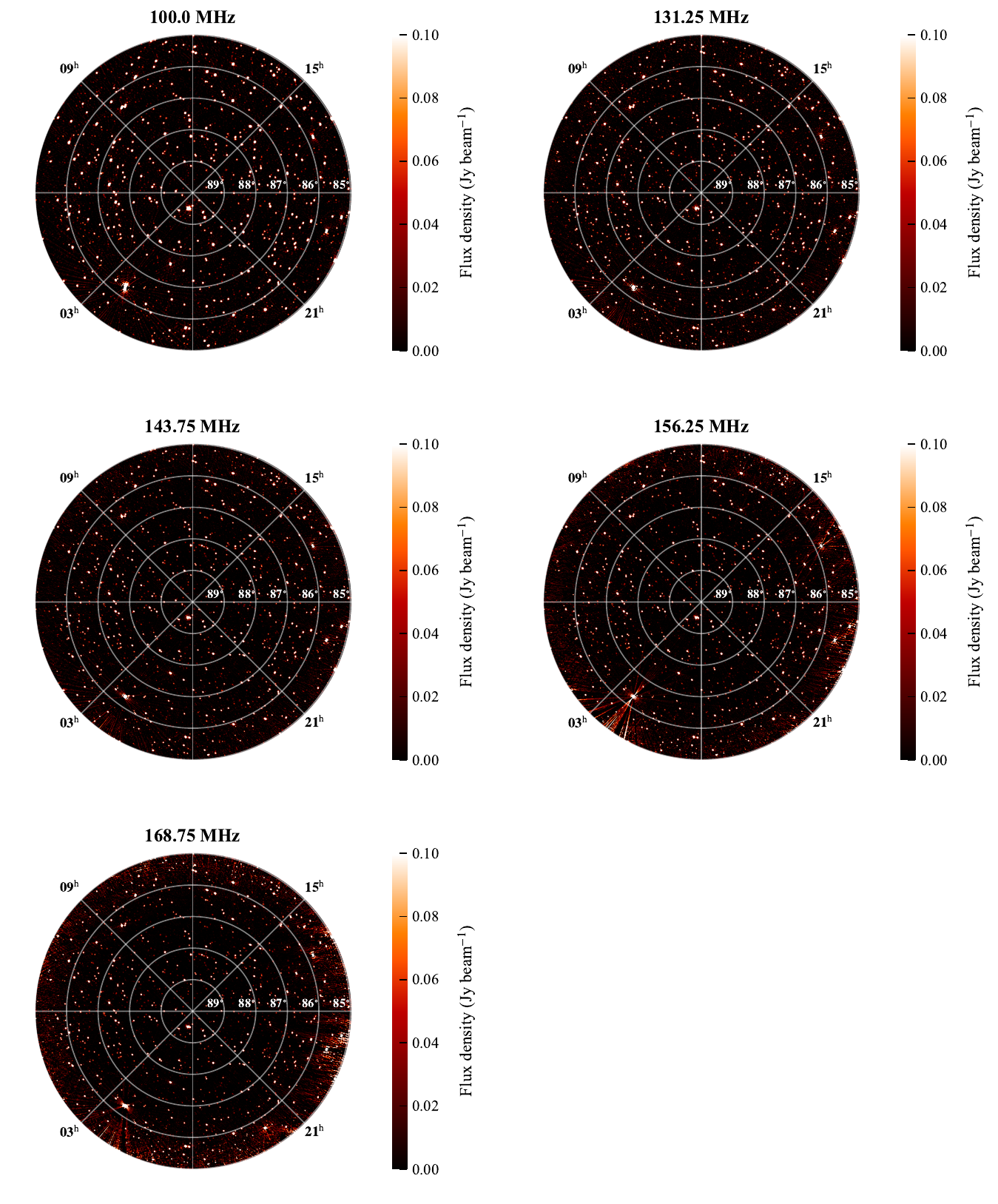}
    \caption{Restored intrinsic-flux images of the NCP region produced with \textsc{DDFacet}. From top to bottom and left to right, the five panels show the final calibrated maps for the five frequency products in order of increasing frequency: 75--125\,MHz (centred at 100\,MHz), 125--137.5\,MHz (131.25\,MHz), 137.5--150\,MHz (143.75\,MHz), 150--162.5\,MHz (156.25\,MHz), and 162.5--175\,MHz (168.75\,MHz). The celestial coordinate grid is overlaid in each panel, with R. A. labelled in hours and Dec. in degrees. An individual colour bar is shown on the right-hand side of each panel, all using the same scale to represent the flux density in units of $\mathrm{Jy\,beam^{-1}}$.}
    \label{fig:ddf_restored_int}
\end{figure*}

Figure \ref{fig:ddf_restored_int} presents the final restored, primary-beam-corrected intrinsic-flux images used for source extraction and catalogue construction for all five 21CMA frequency products, centred at 100, 131.25, 143.75, 156.25, and 168.75\,MHz. To quantify the imaging quality, we calculate the dynamic range (DR), defined as the ratio of the peak flux density in the restored image to the background rms noise, $\mathrm{DR} = S_{\rm peak}/\sigma_{\rm rms}$. The background noise, $\sigma_{\rm rms}$, is measured from the corresponding residual images using sigma-clipped statistics over the target declination range ($\mathrm{Dec} > 85^\circ$). The resulting dynamic ranges are 3081.3, 2556.0, 2519.2, 1800.3, and 1784.0 for the five frequency products, respectively. These values indicate that the direction-dependent calibration and wide-field imaging procedure provides stable high-dynamic-range imaging across the full frequency range. The corresponding residual rms values and restoring-beam parameters are listed in Table~\ref{tab:final_image_properties}.

\begin{table}
    \centering
    \caption{Final image properties. The residual rms values are measured from the corresponding residual images over $\mathrm{Dec}>85^\circ$.}
    \label{tab:final_image_properties}
    \begin{tabular}{cccc}
        \hline
        \shortstack{Frequency\\(MHz)} &
        \shortstack{Restoring beam\\(arcmin)} &
        \shortstack{Residual rms\\(mJy beam$^{-1}$)} &
        \shortstack{Dynamic\\range} \\
        \hline
        100.00 & $2.283\times2.058$ & 8.46 & 3081.3 \\
        131.25 & $1.855\times1.669$ & 9.54 & 2556.0 \\
        143.75 & $1.689\times1.547$ & 9.11 & 2519.2 \\
        156.25 & $1.585\times1.449$ & 11.28 & 1800.3 \\
        168.75 & $1.439\times1.371$ & 11.84 & 1784.0 \\
        \hline
    \end{tabular}
\end{table}

\section{Point Source Identification Strategy}\label{sec:point source finding}

\subsection{Source Finding Method}

With the increasing scale and complexity of radio astronomical data, a variety of automated source-finding algorithms have been developed. Although methods based on watershed segmentation or convolutional neural networks, such as \textsc{ProFound} and \textsc{DeepSource} \citep{Profound, Deepsource}, are increasingly being explored for morphologically complex or diffuse emission, Gaussian-based source finding remains the standard approach for large-scale radio continuum surveys. In particular, \textsc{Aegean} \citep{AEGEAN-1, AEGEAN-2} and \textsc{Selavy} \citep{Selavy} have been widely used in the MWA and ASKAP survey pipelines, respectively.

In this study, we use \textsc{PyBDSF} (Python Blob Detection and Source Finder; \citealt{PyBDSF}) for source extraction and local noise estimation. \textsc{PyBDSF} is widely used in large-scale low-frequency radio surveys, including the LOFAR Two-metre Sky Survey (LoTSS; \citealt{LoTSS-1}). The source-extraction procedure consists of four steps: background mean and rms estimation, island detection, Gaussian fitting, and deconvolution with parameter derivation. Each step is described below.

\subsubsection{Background rms and Mean Estimation}

Because of noise amplification towards the beam edge introduced by primary-beam correction, as well as residuals from the direction-dependent calibration, the image noise is spatially non-uniform. To account for this variation, \textsc{PyBDSF} first estimates the local background mean and rms noise maps using a sliding-box approach. For each pixel $(x, y)$, the local noise, $\sigma_{\rm local}(x, y)$, is measured within a rectangular window centred on that pixel, with iterative $\sigma$-clipping used to reject outliers. This adaptive noise estimation helps suppress spurious detections in regions of elevated noise, such as near the beam edges or around bright sources.

\subsubsection{Island Detection}

Source identification is based on a dual-threshold island strategy. Two thresholds are defined: a pixel threshold, $k_{\rm pix}$, and an island threshold, $k_{\rm isl}$, with $k_{\rm pix} > k_{\rm isl}$. Pixels satisfying $I(x,y) > k_{\rm pix}\,\sigma_{\rm local}(x,y)$ are first identified as candidate peaks. These peaks are then grown into islands by including all contiguous pixels with $I(x,y) > k_{\rm isl}\,\sigma_{\rm local}(x,y)$. The resulting islands represent emission regions that may contain one or more physical sources.

\subsubsection{Gaussian Fitting}

For each detected island, \textsc{PyBDSF} models the image brightness distribution with one or more two-dimensional Gaussian components. The parameters are optimised using a Levenberg--Marquardt non-linear least-squares algorithm. For a single component, the surface brightness model is expressed as
\begin{equation}
    M(x, y) = A \exp \left[ - \frac{1}{2} \left( \frac{x'^2}{\sigma_x^2} + \frac{y'^2}{\sigma_y^2} \right) \right]\,,
\end{equation}
where $A$ is the peak intensity, and $\sigma_x$ and $\sigma_y$ are the Gaussian width parameters along the major and minor axes, respectively. To account for an arbitrary orientation and position, the local coordinates $(x', y')$ are derived from the image coordinates $(x, y)$ through a translation to the component center $(x_0, y_0)$ and a rotation by the position angle $\phi$:
\begin{equation}
    \begin{aligned}
        x' &= (x - x_0) \cos\phi + (y - y_0) \sin\phi \,, \\
        y' &= -(x - x_0) \sin\phi + (y - y_0) \cos\phi \,.
    \end{aligned}
\end{equation}

\subsubsection{Deconvolution and Parameter Derivation}

The fitted Gaussian parameters describe the \textit{observed} source, namely the convolution of the intrinsic source brightness distribution with the restoring beam. To estimate the intrinsic source properties, the beam contribution is deconvolved from the fitted Gaussian parameters.

Assuming an observed source size $\Theta_{\rm obs}$ and a restoring-beam size $\Theta_{\rm beam}$, the intrinsic source size $\Theta_{\rm int}$ can be approximated in terms of the FWHM as
\begin{equation}
    \Theta_{\rm int} \approx \sqrt{\Theta_{\rm obs}^2 - \Theta_{\rm beam}^2}\,.
\end{equation}
If $\Theta_{\rm obs} \approx \Theta_{\rm beam}$, the source is treated as unresolved, and its integrated flux density is taken to be equal to its peak flux density. For resolved sources, the integrated flux density is calculated as
\begin{equation}
    S_{\rm int} = A \frac{\sigma_x \sigma_y}{\sigma_{{\rm beam},x} \sigma_{{\rm beam},y}}\,.
\end{equation}
The final catalogue records the source position (R. A. and Dec.), peak and integrated flux densities, and the deconvolved morphological parameters.

\subsection{Source Extraction Strategy}
\label{subsec:strategy}

To extract radio sources from the 21CMA images at different central frequencies, we use \textsc{PyBDSF}. Because the angular resolution of the array is frequency-dependent ($\theta \propto \lambda / D$), the point spread function (PSF) varies across the observed bands. A fixed rms box is therefore not appropriate for a multi-frequency analysis. To maintain a consistent local noise estimation across the full bandwidth, we adopt an adaptive parameter-setting strategy.

For each input image, we first calculate the geometric mean beam size, $\theta_{\rm beam}$, from the major (\texttt{BMAJ}) and minor (\texttt{BMIN}) axes of the restoring beam recorded in the FITS header:
\begin{equation}
    \theta_{\rm beam} = \sqrt{\texttt{BMAJ} \times \texttt{BMIN}}\,.
\end{equation}
Using the pixel scale, $\Delta_{\rm pix}$, derived from \texttt{CDELT2} (with $|\texttt{CDELT1}| \approx |\texttt{CDELT2}|$ in practice), we then express the beam width in pixels as
\begin{equation}
    N_{\rm beam} = \mathrm{round}\left( \frac{\theta_{\rm beam}}{\Delta_{\rm pix}} \right).
\end{equation}
In \textsc{PyBDSF}, the \texttt{rms\_box} parameter is specified as a pair of values, where the first defines the size of the sliding window used for local background estimation, and the second sets the step size with which the window is moved across the image. To estimate the local noise while suppressing small-scale fluctuations, we set
\begin{equation}
    \texttt{rms\_box} = (5 \times N_{\rm beam},\, N_{\rm beam})\,.
\end{equation}
In addition, to account for PSF variations across the wide-field images, we enable a spatially varying PSF model (\texttt{psf\_vary\_do=True}). For high signal-to-noise regions near exceptionally bright sources ($\mathrm{S/N} > 80$), we also adopt a smaller local rms box, corresponding to three beam widths, so that the local rms estimate can better track small-scale fluctuations associated with imaging artefacts around bright sources.

Artefacts arising from sidelobes or ionospheric residuals can persist near bright sources despite calibration. To minimise false positives and maintain catalogue reliability, we adopted a pixel detection threshold of \texttt{thresh\_pix} = 5.0 ($5\sigma$) and an island boundary threshold of \texttt{thresh\_isl} = 4.0 ($4\sigma$). A $5\sigma$ pixel detection threshold is standard in low-frequency radio continuum surveys \citep[e.g.,][]{MIGHTEE, GMRT2025} as a practical balance between catalogue completeness and the false-detection rate. Furthermore, although the default \textsc{PyBDSF} island threshold is $3\sigma$ \citep{PyBDSF}, we adopted a more conservative $4\sigma$ boundary, following topological flooding strategies similar to those used in the \textsc{Aegean} source finder \citep{AEGEAN-1, AEGEAN-2}. This choice restricts island growth and helps prevent noise spikes or residual sidelobes from being incorporated into source emission islands.

Because of the limited angular resolution of the array, the source finder can occasionally identify residual sidelobes as independent sources or decompose a single physical source into multiple Gaussian components. To assess the reliability of closely separated fitted components, we applied the adjacent-frequency source-validation method of \citet{Zheng2016}. For each frequency product, we identified all component pairs separated by less than the major axis of the corresponding restoring beam, $\theta_{\rm BMAJ}$, and examined each component independently in the adjacent frequency products. A component was retained if a positional counterpart was detected in at least one adjacent product; otherwise, it was excluded as a probable fitting fragment or imaging artefact. Thus, both members of a close pair were retained when each was independently recovered in an adjacent product, whereas only the recurrent component was retained when the other appeared in a single frequency product.

The restoring beam of the 168.75-MHz image is $1.439 \times 1.371$ arcmin, corresponding to an equivalent circular FWHM of approximately 1.4 arcmin. At this resolution, the vast majority of extragalactic radio sources within our target region ($\mathrm{Dec}>85^\circ$) are expected to remain unresolved. This expectation is supported by the source parameters returned by PyBDSF: across the five frequency products, 94.9--98.1 per cent of the detected sources have deconvolved major axes smaller than the major axis of the corresponding restoring beam, while 94.6--98.0 per cent satisfy $S_{\rm int}/S_{\rm peak}<1.5$. These results indicate that the vast majority of the detected sources are unresolved or only marginally resolved. After the automated spatial filtering, we further removed the well-known extended radio galaxy 3C\,61.1 by visual inspection in order to construct a strictly point-source sample. Because of its extended radio lobes, 3C\,61.1 is resolved into multiple substructures by the source finder. After these refinements, the final catalogue is dominated by unresolved or marginally resolved sources, and the subsequent analysis focuses on their positions and total flux densities rather than on detailed morphological information.

\section{Catalogue Construction and Refinement}
\label{sec:rescale}

\subsection{Astrometric Correction}
\label{subsec:astrometry}

After source extraction and morphological filtering, we obtained five independent preliminary catalogues for the sub-bands centred at 100, 131.25, 143.75, 156.25, and 168.75\,MHz. Because the angular resolution of an interferometer scales approximately as $\theta \propto \nu^{-1}$, we adopted the preliminary catalogue at 168.75\,MHz, corresponding to the highest-frequency band, as the astrometric reference. To assess and correct systematic positional offsets arising from ionospheric refraction, instrumental pointing errors, and residual phase-calibration errors, we cross-matched this catalogue against the NRAO VLA Sky Survey (NVSS; \citealp{NVSS}). Observed at 1.4\,GHz with an angular resolution of 45\,arcsec and sub-arcsecond positional accuracy for relatively bright sources, NVSS provides a suitable absolute astrometric reference frame. Using a matching radius of 60\,arcsec, we performed a nearest-neighbour cross-match and calculated the spherical positional offsets in Right Ascension, $\Delta \mathrm{RA}\cos(\mathrm{Dec})$, and Declination, $\Delta \mathrm{Dec}$, for the 21CMA sources relative to their NVSS counterparts.

\begin{figure}
	\centering
	\includegraphics[width=\columnwidth]{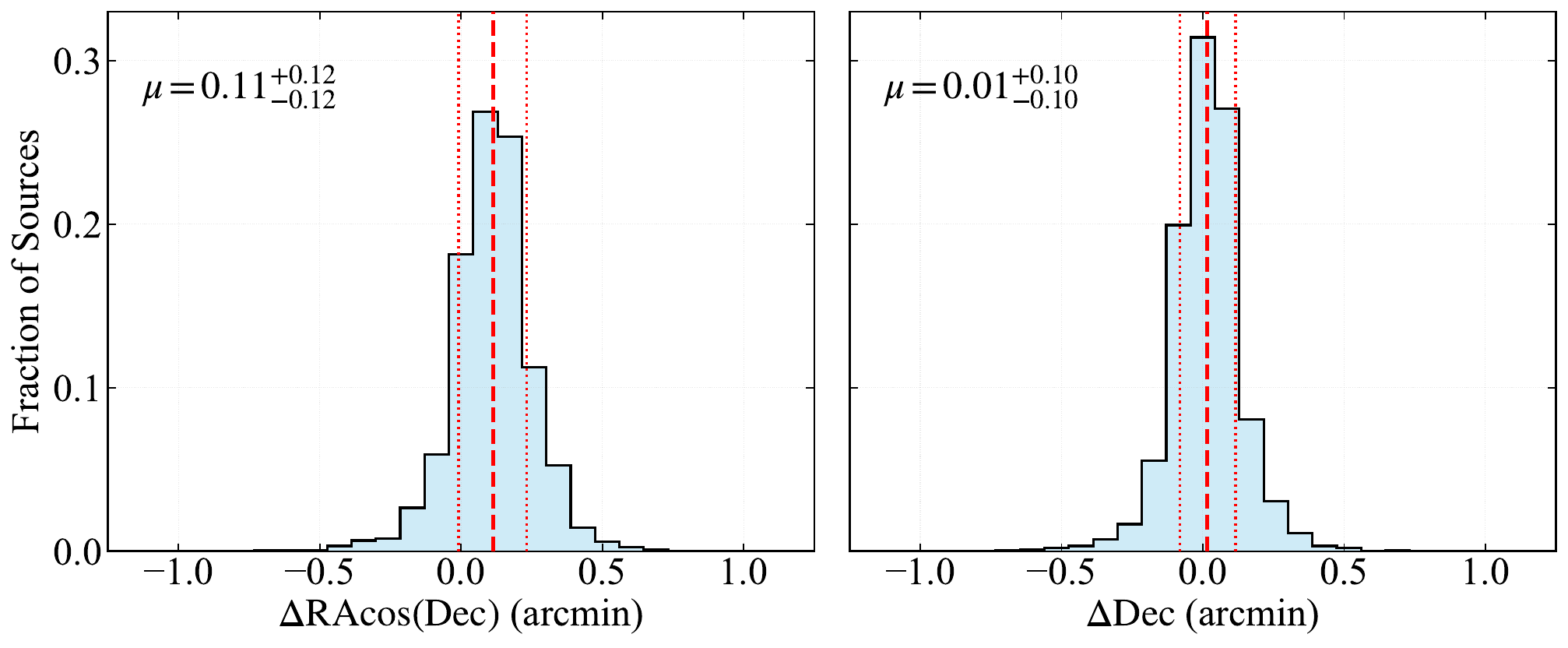}
	\caption{Distributions of the local tangent-plane astrometric residuals between the 21CMA sources and their NVSS counterparts. The left panel shows the east--west component, $\Delta\mathrm{RA}\cos(\mathrm{Dec})$, and the right panel shows the north--south component, $\Delta\mathrm{Dec}$. The mean offset ($\mu$) and the $1\sigma$ dispersion are indicated in the upper-left corner of each panel.}
	\label{fig:astrometry_hist}
\end{figure}

Figure \ref{fig:astrometry_hist} shows the distributions of these astrometric offsets. We find a mean offset of $\Delta \mathrm{Dec} = 0.01 \pm 0.10$\,arcmin in Declination and $\Delta \mathrm{RA}\cos(\mathrm{Dec}) = 0.11 \pm 0.12$\,arcmin in Right Ascension. In Declination, the mean offset of 0.01\,arcmin is smaller than the corresponding $1\sigma$ dispersion of 0.10\,arcmin and is therefore consistent with zero. This indicates that the residuals in Declination are dominated by random positional uncertainties rather than by a systematic astrometric shift. We therefore applied no systematic correction to the Declination coordinates.

In Right Ascension, the mean offset of 0.11\,arcmin is comparable to the corresponding $1\sigma$ dispersion of 0.12\,arcmin. To test whether this offset could arise from uncertainties in the 21CMA sampling timestamps, we estimated the positional drift expected from Earth's rotation within the timestamp error budget. The resulting maximum drift in R. A. is only $\sim 0.005$\,arcmin, well below the observed offset. We therefore conclude that the measured R. A. offset is unlikely to be caused by timing errors and is more plausibly associated with residual instrumental phase systematics or minor errors in the station geodetic coordinates. A uniform correction was therefore applied to the R. A. values of all five preliminary sub-band catalogues after converting the measured offset into the corresponding coordinate shift by dividing by $\cos(\mathrm{Dec})$.

Following the astrometric correction, we performed a global cross-match to merge the five sub-band catalogues. In constructing the preliminary multi-band catalogue, which records the flux densities and associated uncertainties at all five frequencies, we adopted the source position from the highest-frequency available detection for each matched source, since it provides the highest angular resolution. This astrometric correction and multi-band merging procedure yielded a final cross-matched catalogue containing 1746 candidate point sources.

\subsection{Primary Beam Correction and Flux Calibration}
\label{subsec:flux_calibration}

As the 21CMA currently lacks an empirically calibrated primary-beam model derived from dedicated observations, owing to the technical difficulty of direct beam measurements for a large low-frequency array, we relied on the theoretical beam model obtained via electromagnetic simulations for initial imaging and calibration. However, discrepancies between the theoretical model and the actual instrumental response introduce direction-dependent systematic errors in the extracted intrinsic flux densities. To reduce these systematics and improve the flux-density calibration, we derived an empirical correction that accounts for both a global frequency-dependent flux-scale offset and a residual radial variation.

To derive the flux-correction functions, we extracted a calibration sample from the preliminary catalogues of the five frequency products by requiring a peak signal-to-noise ratio of $\mathrm{S/N} > 7$, where $\mathrm{S/N} = S_{\rm peak}/\sigma_{\rm local}$, and an integrated flux density of $S_{\rm int}>0.05$\,Jy. Because morphological spatial filtering had already been applied to remove obvious artefacts, this threshold retains a sufficiently large fitting sample while excluding most low-S/N fragments. In the subsequent analysis, we used only the integrated flux densities and their associated uncertainties.

The calibration sample was cross-matched with the 144\,MHz LoTSS-DR3 catalogue \citep{LoTSS-DR3} using a matching radius of 60\,arcsec. To reduce discrepancies arising from the different angular resolutions and source-component definitions of the two surveys, we retained only LoTSS-DR3 sources classified as single-component sources (\texttt{S}) with $S_{\rm int}/S_{\rm peak}<1.5$ and integrated flux densities greater than 0.05\,Jy. Duplicate associations were removed so that each catalogue source contributed to no more than one matched pair. To estimate the reference flux density, $S_{\rm ref}$, at the central frequencies of the five 21CMA frequency products, we adopted a representative spectral index derived from sources jointly matched among LoTSS-DR3 \citep{LoTSS-DR3}, WENSS \citep{WENSS}, and NVSS \citep{NVSS} within the target region ($\mathrm{Dec} > 85^\circ$). Prior to the spectral fitting, both the WENSS flux densities and their associated uncertainties were scaled by a factor of 0.9 to align them with the adopted low-frequency flux-density scale \citep{WENSS-ratio}. The resulting median spectral index was $\alpha=-0.712$, which was used to extrapolate the LoTSS-DR3 flux densities according to $S_{\rm ref}(\nu)=S_{\rm LoTSS,144}(\nu/144\,\mathrm{MHz})^{-0.712}$. We then calculated the flux-density ratio, $R=S_{\rm obs}/S_{\rm ref}$, for each matched source.

The flux-density discrepancies can include both a global frequency-dependent scale offset and a radial component associated with residual inaccuracies in the theoretical primary-beam model. We therefore analysed the flux-density ratio, $R$, as a function of the angular distance $\theta$ from the NCP, where $\theta=90^\circ-\mathrm{Dec}$. The binned median ratios were fitted with
\begin{equation}
    R(\theta)=R_0\exp\left[B\left(\frac{\theta}{\theta_{\rm HP}}\right)^n\right]\, .
\end{equation}
Here, $\theta_{\rm HP}$ is the frequency-dependent primary-beam half-power radius, $R_0$ represents the global flux-density-scale normalisation, and the exponential term describes the residual radial variation. The parameters $R_0$, $B$, and $n$ were fitted independently for the five frequency products. The \textsc{NEC2++} model is applied as a common station-beam model across the array; the empirical correction therefore characterises the array-averaged and time-averaged residual response rather than station-specific beam variations.

\begin{figure*}
	\centering
	\includegraphics[width=2\columnwidth]{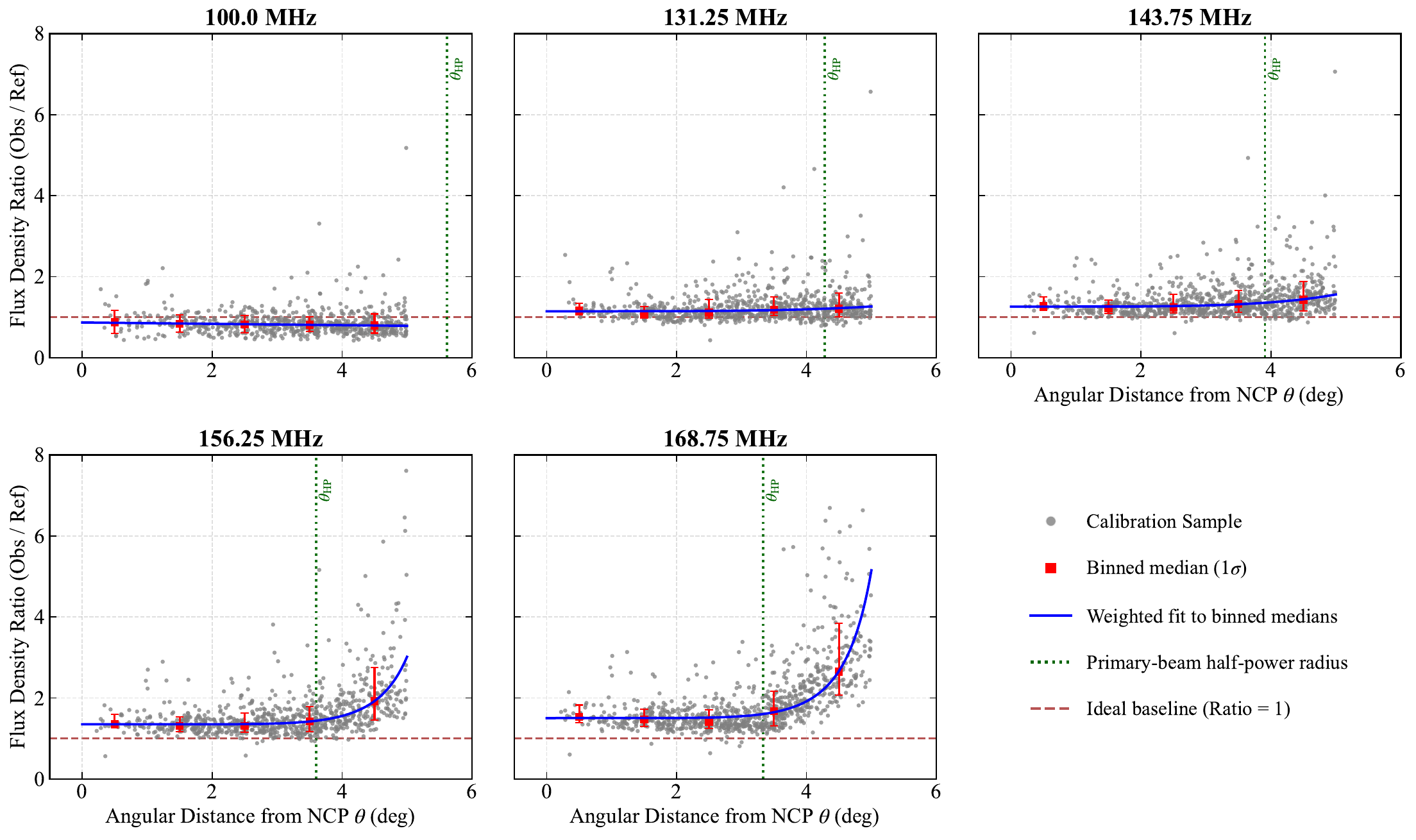}
	\caption{Distributions of the measured-to-reference flux-density ratios ($S_{\rm obs}/S_{\rm ref}$) as a function of angular distance from the NCP ($\theta=90^\circ-\mathrm{Dec}$) for the five 21CMA frequency products. The reference flux densities were extrapolated from the LoTSS-DR3 144\,MHz catalogue using a median spectral index of $\alpha=-0.712$ derived for the target region. Grey dots represent the selected calibration sources, while red squares and their error bars denote the binned median ratios and their $1\sigma$ dispersions. The solid blue curves show the weighted fits of $R(\theta)=R_0\exp[B(\theta/\theta_{\rm HP})^n]$ to the binned medians. The green dotted vertical line in each panel marks the frequency-dependent primary-beam half-power radius, $\theta_{\rm HP}$, and the red dashed horizontal line marks the ideal ratio $R=1$.}
	\label{fig:flux_ratio_fit}
\end{figure*}

Figure \ref{fig:flux_ratio_fit} shows the scatter distributions, binned medians, and fitted correction functions for the five sub-bands. The flux-density ratios and their scatter increase near and beyond the half-power radius, particularly at 156.25 and 168.75\,MHz. This suggests that, although the simulated electromagnetic beam model improves upon a simple Gaussian approximation, residual mismatches with the true beam response remain in the outer part of the main beam. As the beam becomes narrower at higher frequencies, these differences between the model and the actual response become more apparent towards the edge of the NCP field.

The derived continuous correction functions were applied globally to the cross-matched master catalogue containing 1746 candidate point sources. For each source, the angular distance $\theta$ was calculated and substituted into the corresponding correction function to obtain the correction factor, $R_\nu(\theta)$. Both the measured integrated flux densities and their uncertainties were then divided by this factor to complete the empirical direction-dependent flux-density rescaling.

\subsection{Source Selection and Spectral Fitting}
\label{subsec:spectral_fitting}

Following the multi-band source extraction and direction-dependent flux calibration, our preliminary master catalogue spans the frequency range 75--175\,MHz. For extragalactic radio sources, emission in this low-frequency regime is predominantly governed by synchrotron radiation and is typically well described by a power-law spectrum, $S \propto \nu^{\alpha}$. To characterise the spectral properties of these sources and to further reject spurious detections or noise peaks, we derived an in-band spectral index, $\alpha_{\rm in\text{-}band}$, for each candidate source across the 21CMA observing band.

To derive the in-band spectral index, we performed a weighted fit in natural logarithmic space with a first-order power-law model. Adopting a reference frequency of $\nu_0 = 100$\,MHz, the model is written as
\begin{equation}
    \ln S(\nu) = \alpha_{\rm in\text{-}band} \ln\left(\frac{\nu}{\nu_0}\right) + c_0 \, ,
\end{equation}
where $S(\nu)$ is the flux density at frequency $\nu$, and $c_0$ is the logarithmic normalisation constant. The individual flux-density uncertainties, $\sigma_S$, were propagated into logarithmic space as $\sigma_{\ln S} = \sigma_S / S$ and used as the fitting weights.

To ensure the reliability of the spectral fitting and the physical credibility of the extracted sources, we imposed a minimum frequency-coverage requirement. Specifically, the fit was restricted to sources with independent valid detections in at least three of the five 21CMA sub-bands. Candidates that did not satisfy this criterion were regarded as likely unreliable and were excluded from the master catalogue.

After applying this frequency-coverage filter, and excluding a negligible number of additional outliers for which the fitting failed to converge, we obtained in-band spectral indices and their associated fitting uncertainties, derived from the diagonal elements of the covariance matrix. This multi-band selection criterion yielded a highly reliable point-source catalogue for the  core region, containing a final sample of 1246 radio sources.

\section{Catalogue Overview}
\label{sec:catalogue}

We present the 21CMA-DDR1 catalogue for the NCP region, where DDR1 denotes the first direction-dependent data release of the 21CMA point-source catalogue. The catalogue is based on a continuous 24-h observation obtained on 2013 March 25 and covers the central field defined by $\mathrm{Dec} > 85^\circ$, corresponding to an area of approximately 78\,deg$^2$. After morphological filtering, direction-dependent flux calibration, and in-band spectral fitting, the final catalogue contains 1246 extragalactic radio sources.

For each source, the catalogue provides astrometric positions, systematically aligned to the NVSS reference frame, and integrated flux densities measured in the five 21CMA sub-bands, namely 75--125, 125--137.5, 137.5--150, 150--162.5, and 162.5--175\,MHz, centred at 100, 131.25, 143.75, 156.25, and 168.75\,MHz, respectively, together with their $1\sigma$ uncertainties. The quoted $1\sigma$ uncertainties are the formal integrated-flux-density uncertainties returned by \textsc{PyBDSF}; during the empirical flux-density rescaling, they were divided by the same correction factor as the corresponding integrated flux densities. These formal uncertainties do not include the residual flux-scale uncertainty estimated in Section~\ref{sec:flux_val}. 
The catalogue also includes the in-band spectral index, $\alpha$, and its fitting uncertainty derived from linear regression in logarithmic space.

A representative sample of the catalogue is shown in Table \ref{tab:catalogue_sample}. For illustration, we list 10 representative sources to demonstrate the data format. The full machine-readable catalogue, containing all 1246 sources and 16 data columns, is available in the online supplementary material. A detailed description of the catalogue columns is provided in Appendix \ref{appendix:catalogue_columns}.

\begin{table*}
    \centering
    \caption{A representative sample of 10 sources from the 21CMA-DDR1 catalogue. Flux densities and spectral indices are given as value $\pm$ $1\sigma$.}
    \label{tab:catalogue_sample}
    \resizebox{\textwidth}{!}{
    \begin{tabular}{lccccccccc}
        \hline
        Source\_ID &
        R.A. &
        Dec. &
        Type &
        $S_{100}$ &
        $S_{131.25}$ &
        $S_{143.75}$ &
        $S_{156.25}$ &
        $S_{168.75}$ &
        $\alpha$ \\
        &
        (J2000) &
        (J2000) &
        &
        (Jy) &
        (Jy) &
        (Jy) &
        (Jy) &
        (Jy) &
        \\
        \hline

        21CMA-DDR1-J163123+893352 &
        16:31:23.34 &
        +89:33:51.9 &
        P &
        $0.240 \pm 0.011$ &
        $0.190 \pm 0.008$ &
        $0.187 \pm 0.006$ &
        $0.183 \pm 0.006$ &
        $0.166 \pm 0.004$ &
        $-0.662 \pm 0.092$ \\

        21CMA-DDR1-J023440+891718 &
        02:34:39.67 &
        +89:17:18.0 &
        P &
        $0.676 \pm 0.027$ &
        $0.524 \pm 0.017$ &
        $0.476 \pm 0.016$ &
        $0.460 \pm 0.018$ &
        $0.432 \pm 0.010$ &
        $-0.840 \pm 0.085$ \\

        21CMA-DDR1-J114013+890915 &
        11:40:13.39 &
        +89:09:15.1 &
        P &
        $0.389 \pm 0.019$ &
        $0.272 \pm 0.010$ &
        $0.264 \pm 0.012$ &
        $0.249 \pm 0.008$ &
        $0.235 \pm 0.006$ &
        $-0.886 \pm 0.099$ \\

        21CMA-DDR1-J175754+890654 &
        17:57:53.82 &
        +89:06:53.8 &
        P &
        $0.351 \pm 0.013$ &
        $0.245 \pm 0.007$ &
        $0.238 \pm 0.006$ &
        $0.236 \pm 0.005$ &
        $0.220 \pm 0.004$ &
        $-0.791 \pm 0.070$ \\

        21CMA-DDR1-J055903+890450 &
        05:59:02.61 &
        +89:04:49.8 &
        P &
        $0.322 \pm 0.013$ &
        $0.281 \pm 0.010$ &
        $0.267 \pm 0.008$ &
        $0.248 \pm 0.007$ &
        $0.236 \pm 0.005$ &
        $-0.613 \pm 0.079$ \\

        21CMA-DDR1-J125555+885926 &
        12:55:55.09 &
        +88:59:26.0 &
        P &
        $0.447 \pm 0.017$ &
        $0.317 \pm 0.011$ &
        $0.302 \pm 0.010$ &
        $0.292 \pm 0.010$ &
        $0.279 \pm 0.006$ &
        $-0.849 \pm 0.081$ \\

        21CMA-DDR1-J231025+885721 &
        23:10:24.61 &
        +88:57:21.1 &
        P &
        $0.301 \pm 0.014$ &
        $0.223 \pm 0.011$ &
        $0.208 \pm 0.008$ &
        $0.202 \pm 0.007$ &
        $0.202 \pm 0.005$ &
        $-0.728 \pm 0.098$ \\

        21CMA-DDR1-J050103+884832 &
        05:01:02.64 &
        +88:48:32.4 &
        P &
        $1.019 \pm 0.023$ &
        $0.785 \pm 0.016$ &
        $0.740 \pm 0.011$ &
        $0.704 \pm 0.011$ &
        $0.668 \pm 0.008$ &
        $-0.782 \pm 0.047$ \\

        21CMA-DDR1-J141116+884754 &
        14:11:16.08 &
        +88:47:54.0 &
        P &
        $0.409 \pm 0.017$ &
        $0.296 \pm 0.009$ &
        $0.282 \pm 0.007$ &
        $0.279 \pm 0.007$ &
        $0.272 \pm 0.004$ &
        $-0.653 \pm 0.076$ \\

        21CMA-DDR1-J134923+884638 &
        13:49:22.61 &
        +88:46:37.8 &
        P &
        $0.709 \pm 0.018$ &
        $0.513 \pm 0.011$ &
        $0.485 \pm 0.009$ &
        $0.466 \pm 0.009$ &
        $0.439 \pm 0.007$ &
        $-0.870 \pm 0.054$ \\

        \hline
    \end{tabular}
    }
\end{table*}

\section{Catalogue Validation}
\label{sec:validation}

\subsection{Astrometric Accuracy}
\label{sec:astrometry_val}

To evaluate the astrometric precision of the final 21CMA catalogue, we cross-matched the 1246 point sources with the 1.4\,GHz NVSS catalogue and obtained 1230 successful matches. Given its high positional accuracy in the NCP region, NVSS serves as the astrometric reference. For the matched sources, we calculated the positional residuals in R. A. and Dec. To account for projection effects at high declination, the R. A. offsets were weighted by $\cos(\mathrm{Dec})$:
\begin{equation}
\begin{array}{lcl}
\Delta \mathrm{RA}_{\rm proj} & = & (\mathrm{RA}_{\rm 21CMA} - \mathrm{RA}_{\rm NVSS}) \cos(\mathrm{Dec}) \,, \\
\Delta \mathrm{Dec}           & = & \mathrm{Dec}_{\rm 21CMA} - \mathrm{Dec}_{\rm NVSS} \, .
\end{array}
\end{equation}

\begin{figure}
	\centering
	\includegraphics[width=\columnwidth]{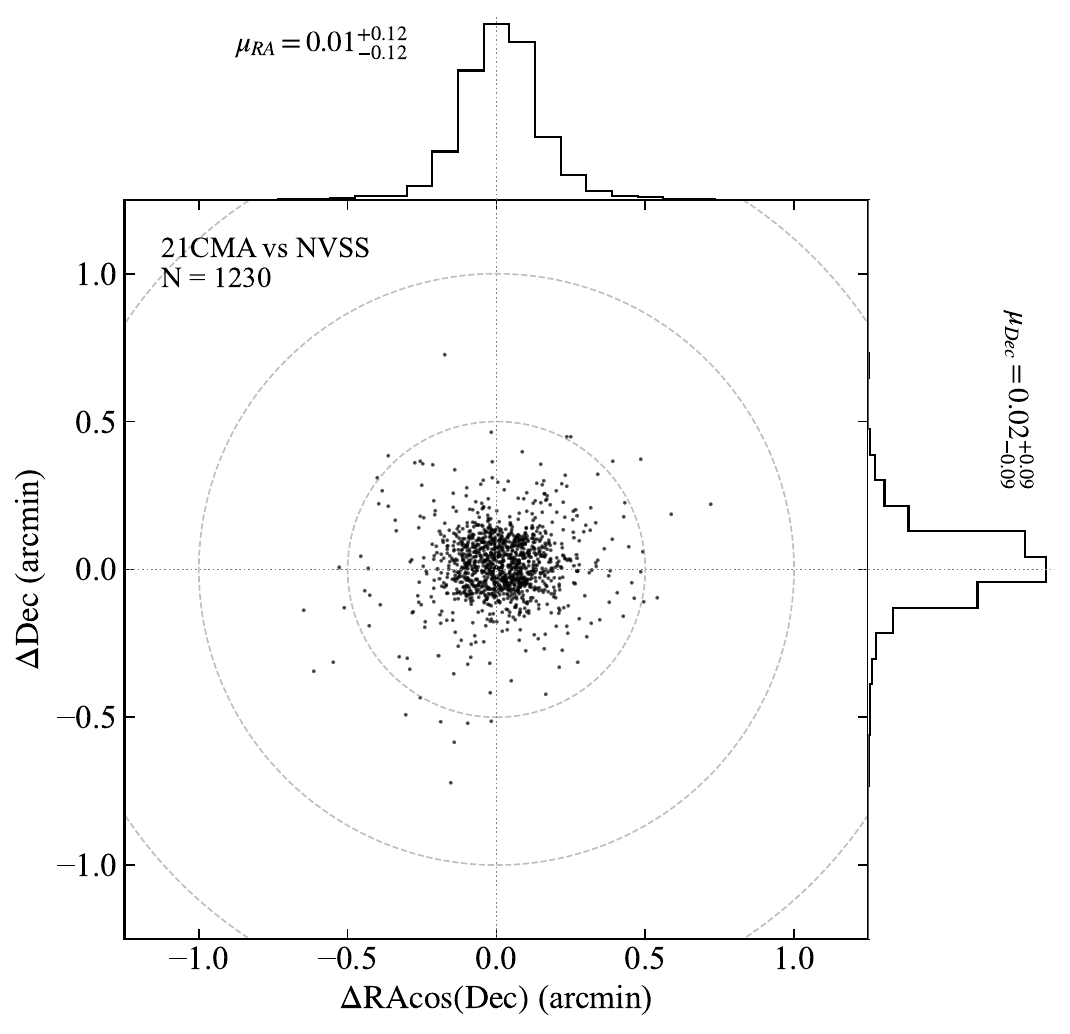}
	\caption{Astrometric accuracy of the 21CMA catalogue derived from the comparison with the NVSS catalogue. The scatter plot presents the positional offsets in Right Ascension ($\Delta \mathrm{RA} \cos(\mathrm{Dec})$) and Declination ($\Delta \mathrm{Dec}$) for the 1230 matched sources. The concentric dashed circles indicate radial offsets in increments of $0.5$\,arcmin. The top and right panels display the marginalised histograms for the R. A. and Dec. residuals, respectively. The mean offsets ($\mu$) and their corresponding $1\sigma$ dispersions are annotated in arcminutes.}
	\label{fig:astrometry_val}
\end{figure}

Figure \ref{fig:astrometry_val} shows the distributions of these astrometric residuals. We find mean offsets of $0.01 \pm 0.12$\,arcmin in R. A. and $0.02 \pm 0.09$\,arcmin in Dec. Both residuals are statistically consistent with zero, confirming the effectiveness of the astrometric correction applied during the data reduction.

In previous NCP catalogues based on 21CMA observations, \citet{Zheng2016} and \citet{Zhao2022} reported typical positional accuracies of $\sim 5\arcmin$ and $\sim 2\arcmin$, respectively. By comparison, the $1\sigma$ dispersion of the positional residuals in the present catalogue is only $\sim 0.1\arcmin$, representing an improvement of more than an order of magnitude. This improvement reflects the effectiveness of the direction-dependent calibration and astrometric alignment procedures adopted in this work. Moreover, the residual dispersion is much smaller than the synthesised beam size of the 21CMA images in this frequency range. The final catalogue therefore localises sources to well within the angular resolution of the array.

\subsection{Flux Density Consistency}
\label{sec:flux_val}

To assess the photometric consistency of the 21CMA catalogue over the 75--175\,MHz band, we compared the corrected integrated flux densities of the 21CMA sources with those reported in the LoTSS-DR3 catalogue at 144\,MHz \citep{LoTSS-DR3}. Because LoTSS-DR3 covers the same North Celestial Pole region at a closely comparable frequency, it provides a useful reference for evaluating the residual consistency of the corrected flux-density scale.

For a direct comparison, we first extrapolated the LoTSS-DR3 reference flux densities to the central frequencies of the five 21CMA sub-bands. Consistent with the value derived in the DD flux-calibration procedure (Section \ref{subsec:flux_calibration}), we adopted a local median spectral index of $\alpha=-0.712$ for this extrapolation. We then calculated the flux-density ratio $S_{\rm 21CMA}/S_{\rm LoTSS,scaled}$ at each central frequency.

\begin{figure}
	\centering
	\includegraphics[width=\columnwidth]{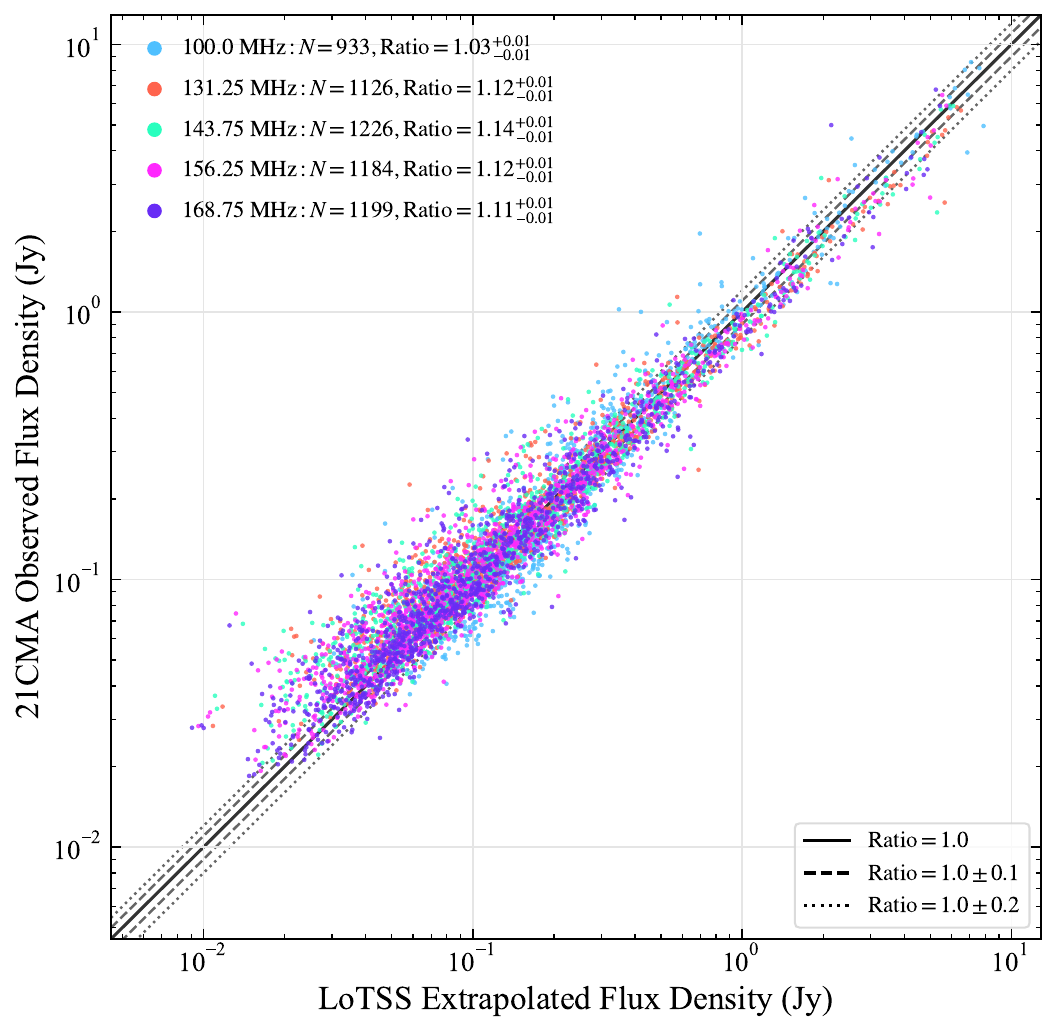}
	\caption{Comparison of the corrected 21CMA integrated flux densities with the extrapolated LoTSS-DR3 reference flux densities at the five central frequencies from 100 to 168.75\,MHz. The 144\,MHz LoTSS-DR3 flux densities were scaled to the corresponding observing frequencies using the median spectral index $\alpha=-0.712$. The solid black line indicates the one-to-one relation, while the dashed and dotted lines mark the $\pm10$ and $\pm20$ per cent deviation boundaries, respectively. The legend gives the number of matched sources and the median flux-density ratio with its $1\sigma$ standard error for each sub-band.}
	\label{fig:flux_val}
\end{figure}

Figure \ref{fig:flux_val} shows the scatter distribution of the 21CMA measured flux densities against the extrapolated LoTSS-DR3 reference fluxes. The median flux-density ratios for the full matched samples are $1.03\pm0.01$ at 100\,MHz, $1.12\pm0.01$ at 131.25\,MHz, $1.14\pm0.01$ at 143.75\,MHz, $1.12\pm0.01$ at 156.25\,MHz, and $1.11\pm0.01$ at 168.75\,MHz.

Despite the inherent differences in instrumental architecture, angular resolution, and ionospheric conditions between the two arrays, the flux-density ratios show reasonable overall agreement with unity. This indicates that the data-reduction pipeline largely corrects for the primary-beam attenuation and provides a broadly consistent flux-density scale across the 75--175\,MHz band.

To estimate the residual band-dependent flux-scale uncertainty, we selected matched sources with an integrated-flux-density signal-to-noise ratio of $\mathrm{S/N}\geq15$. For each frequency band $\nu$, the reference flux density of each matched source $i$ was obtained by extrapolating its LoTSS-DR3 144-MHz flux density as
$S_{\nu,\mathrm{ref},i}=S_{144,\mathrm{LoTSS},i}(\nu/144\,\mathrm{MHz})^{-0.712}$.
We then defined the source-by-source flux-density ratio as
$R_{\nu,i}=S_{\nu,\mathrm{21CMA},i}/S_{\nu,\mathrm{ref},i}$.
Denoting the median ratio in each frequency band by
$\widetilde{R}_{\nu}=\operatorname{median}_{i}(R_{\nu,i})$,
we estimated the fractional uncertainty from the normalised median absolute deviation \citep{RousseeuwCroux1993},
\begin{equation}
f_{\rm cal,\nu}
=
\frac{
1.4826\,
\operatorname{median}_{i}
\left(
\left|R_{\nu,i}-\widetilde{R}_{\nu}\right|
\right)
}{
\widetilde{R}_{\nu}
}.
\label{eq:external_flux_uncertainty}
\end{equation}

\begin{table}
    \centering
    \caption{Empirical residual flux-scale uncertainties derived from the high-S/N comparison with LoTSS-DR3.}
    \label{tab:flux_calibration_uncertainty}
    \begin{tabular}{cccc}
        \hline
        \shortstack{Frequency\\(MHz)} &
        \shortstack{Number of\\sources} &
        \shortstack{Median flux\\density ratio} &
        \shortstack{$f_{\rm cal,\nu}$\\(per cent)} \\
        \hline
        100.00 & 295 & 1.063 & 16.9 \\
        131.25 & 333 & 0.933 & 14.4 \\
        143.75 & 381 & 0.948 & 14.7 \\
        156.25 & 333 & 0.946 & 14.7 \\
        168.75 & 427 & 0.949 & 17.2 \\
        \hline
    \end{tabular}
\end{table}

The empirical residual flux-scale uncertainties listed in Table~\ref{tab:flux_calibration_uncertainty} range from 14.4 to 17.2 per cent. These values describe the robust scatter relative to the LoTSS-DR3 reference scale and may also include contributions from spectral extrapolation, differences in angular resolution, and catalogue matching, rather than representing purely instrumental calibration errors.

\subsection{Reliability}
\label{sec:reliability}

To evaluate the reliability of the 21CMA catalogue, defined here as the probability that a detected source corresponds to a genuine astrophysical object rather than instrumental noise or a processing artefact, we adopted a dual-validation strategy. This consists of an empirical cross-match with external surveys together with a statistical false detection rate (FDR; \citealp{FDR}) analysis based on negative-image testing \citep{GMRT2025}.

First, genuine extragalactic radio sources are generally expected to have counterparts in external radio surveys. We therefore cross-matched our final catalogue of 1246 sources with three independent surveys: LoTSS-DR3 (144\,MHz), WENSS (325\,MHz), and NVSS (1.4\,GHz), using a matching radius of 60\,arcsec. We found that 1245 sources, corresponding to 99.92\% of the catalogue, have high-confidence counterparts in at least one of these surveys. This high cross-matching rate supports the reliability of the final catalogue. The remaining unmatched source (0.08\%) satisfies the internal multi-band detection criteria, but its nominal in-band spectral index of $\alpha=-2.27\pm1.66$ is poorly constrained and does not permit a firm spectral classification.

Second, to quantify the background FDR statistically from the image properties, we performed a standard negative-map test. For each sub-band, we generated a negative image by multiplying the calibrated restored image by $-1$. Blind source extraction was then carried out on these negative images using the same \textsc{PyBDSF} configuration as for the positive images, including the $5\sigma$ pixel detection threshold and the spatially varying PSF model.

The results show that no false positive sources were extracted from the negative maps at any of the five central frequencies (100, 131.25, 143.75, 156.25, and 168.75\,MHz), implying no statistically detected false positives in this test. This result is primarily attributable to the effectiveness of the direction-dependent calibration adopted in this work, which substantially improves the image quality by suppressing residual sidelobes and non-Gaussian background structure. In addition, the strict $5\sigma$ detection threshold and the automated spatial filtering provide strong protection against random background fluctuations. Taken together, these tests indicate that the data-reduction and source-extraction pipeline achieves very high reliability across the observing band, with negligible contamination of the final catalogue by spurious artefacts.

\subsection{Comparison with Previous 21CMA Work}
\label{sec:catalogue_comparison}

To assess the impact of the upgraded data-reduction and imaging pipeline, we cross-matched the 21CMA-DDR1 catalogue with the previous 21CMA NCP catalogue published by \citet{Zheng2016} (hereafter Z2016). The Z2016 catalogue was derived from comparable 21CMA observations and covers nearly the same region ($\mathrm{Dec} > 85^\circ$). Given the approximately 4\,arcmin angular resolution of the Z2016 images and the higher angular resolution of the present work, which may resolve some extended sources into multiple components, we adopted a matching radius of 210\,arcsec (approximately 3.5\arcmin). In cases of multiple matches, only the nearest counterpart was retained.

\begin{figure*}
    \centering
    \includegraphics[width=2\columnwidth]{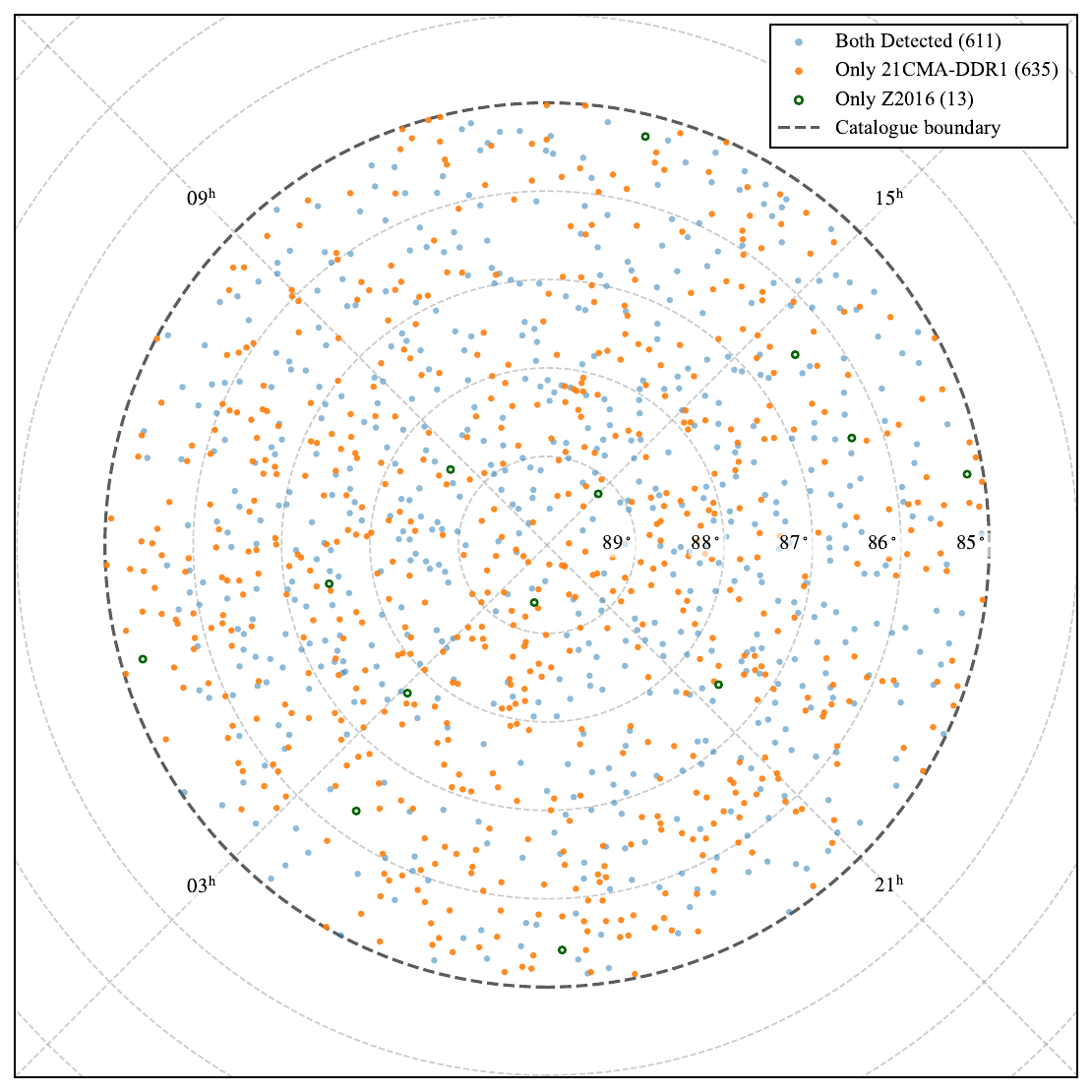}
    \caption{Comparison of the spatial distribution of radio sources in the NCP region ($\mathrm{Dec} > 85^\circ$) between the 21CMA-DDR1 catalogue and the Z2016 catalogue \citep{Zheng2016}. Blue dots represent the 611 sources detected in both catalogues. Open dark green circles mark the 13 sources appearing only in Z2016, while orange dots indicate the 635 sources newly identified in 21CMA-DDR1.}
    \label{fig:sky_comparison}
\end{figure*}

The spatial distribution of the matched sources is shown in Figure \ref{fig:sky_comparison}. Within the field defined by $\mathrm{Dec} > 85^\circ$, the two catalogues share 611 common radio sources. Z2016 contains 13 unmatched sources, which may in part reflect residual RFI or artefacts associated with ionospheric phase fluctuations in the earlier processing. By contrast, our new catalogue contains 635 additional unmatched sources, yielding a total of 1246 sources, approximately twice the number reported in Z2016.

Beyond the algorithmic updates, the difference in sample composition is also influenced by the catalogue-construction strategy. Z2016 includes several resolved sources, most notably the bright radio galaxy 3C\,61.1. By contrast, the present work is designed to construct a high-purity point-source sample, and such extended sources were therefore excluded during the source-extraction stage through morphological filtering (see Section \ref{subsec:strategy}).

The substantial increase in source count primarily reflects the improvement in image fidelity and effective depth achieved with the updated data-reduction strategy. Earlier processing relied on DI calibration and conventional imaging, and was therefore more susceptible to residual ionospheric phase distortions and strong sidelobe artefacts around bright sources, particularly in the outer parts of the field where the calibration quality was less uniform. These residual phase errors and artefacts not only produced spurious structures around bright sources, but also enhanced the local background fluctuations and reduced the contrast of faint compact sources, thereby lowering the effective detection depth of the catalogue. In the present work, the introduction of DDC, together with a more suitable directional partitioning of the field and the corresponding assignment of calibration sources, substantially suppresses these residual artefacts and improves the local calibration quality across the image. As a result, the updated processing recovers a population of faint radio sources that was previously difficult to detect because of dynamic-range limitations.

\section{Discussion}\label{sec:discussion}

\subsection{Spectral Index Comparisons}
\label{subsec:spectral_index}

To characterise the spectral properties of the radio source population detected by the 21CMA, we analysed their spectral energy distributions (SEDs). At frequencies around $\sim 150\,\mathrm{MHz}$, the emission from extragalactic radio sources is typically dominated by synchrotron radiation, and is therefore often described by a power-law spectrum, $S \propto \nu^\alpha$, with a characteristic spectral index of $\alpha \simeq -0.7$. To test this expectation observationally and to assess possible spectral differences within the sample, we characterised the source spectra in two complementary ways, using broad-band spectral indices derived from external surveys and in-band spectral indices measured solely within the 21CMA frequency coverage.

To extend the spectral analysis beyond the 21CMA band, we incorporated flux-density measurements from three external surveys, LoTSS-DR3 (144\,MHz), WENSS (325\,MHz), and NVSS (1.4\,GHz). Using the cross-matched survey measurements, we derived broad-band spectral indices using the external data, while the in-band spectral index, $\alpha_{\rm in\text{-}band}$, was determined independently from the 21CMA sub-bands alone. The main advantage of the in-band spectral index is that it is derived entirely within a single instrument and data-reduction framework, thereby reducing systematic uncertainties associated with cross-survey flux-scale differences and long-term source variability. To ensure statistical robustness, we required a spectral-index uncertainty of $\sigma_\alpha < 0.4$.

\begin{figure}
    \centering
    \includegraphics[width=\columnwidth]{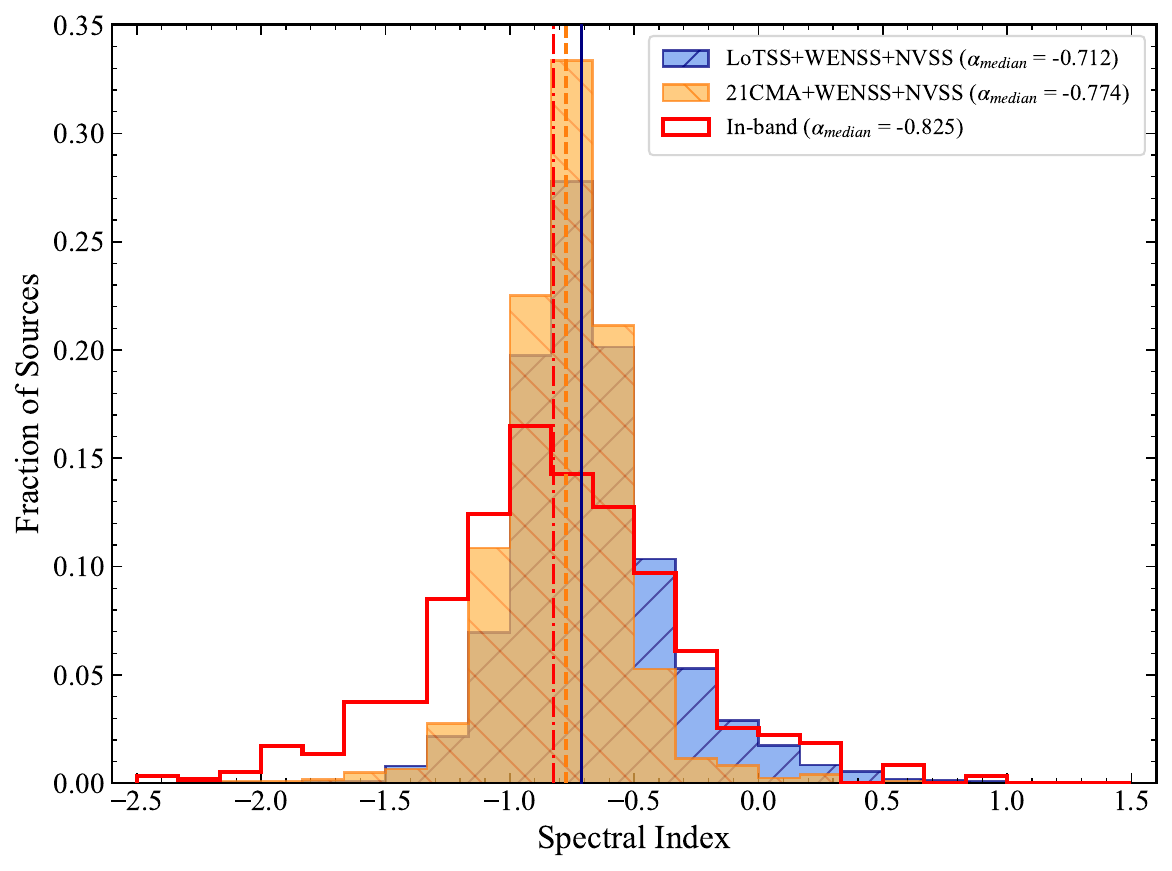}
    \caption{Normalised distributions of the spectral index $\alpha$ for three different data combinations. The blue hatched histogram represents the sample with spectral indices fitted using only external surveys (LoTSS, WENSS, and NVSS), while the orange hatched histogram shows the joint fit combining 21CMA data with WENSS and NVSS. The red step histogram denotes the in-band sample fitted exclusively with the 21CMA sub-bands. The vertical lines mark the median values of the three samples. To ensure the reliability of the fits, only sources with a fitting uncertainty of $\sigma_\alpha < 0.4$ are included.}
    \label{fig:alpha_dist}
\end{figure}

Figure \ref{fig:alpha_dist} shows the normalised distributions of the spectral indices for three data combinations: external surveys only (LoTSS+WENSS+NVSS; 2001 sources), 21CMA combined with external surveys (21CMA+WENSS+NVSS; 1235 sources), and 21CMA in-band only (588 sources). The three distributions broadly overlap within the expected synchrotron-emission regime, with median values of $-0.712$, $-0.774$, and $-0.825$, respectively. The 21CMA in-band median is consistent with spectral-index
distributions reported for other low-frequency-selected radio-source samples. In particular, the GLEAM survey derived in-band spectral indices over 72--231\,MHz and found median values of $\alpha=-0.78$, $-0.79$, $-0.83$, and $-0.83$ in four successive flux-density bins defined at 200\,MHz, with the two brighter samples ($S_{200}\geq0.5$\,Jy) both having a median of
$\alpha=-0.83$ \citep{GLEAM-1}. Broad-band measurements of
low-frequency-selected samples yield comparable values; for example, the LOFAR Bo\"otes survey found a median spectral index of $\alpha=-0.79\pm0.01$ between 150\,MHz and 1.4\,GHz \citep{e-curve-bootes}, while previous 150-MHz studies have reported typical median values spanning approximately $-0.78$ to $-0.87$ \citep{Williams2013,Intema2011,IshwaraChandra2010,Sirothia2009, IshwaraChandraMarathe2007}. The somewhat broader distribution of the in-band spectral indices (red histogram) is also expected, because the 21CMA observing band (75--175\,MHz) spans a relatively limited range in logarithmic frequency space and therefore provides a shorter fitting lever arm. As a result, the fitted slope is more sensitive to the flux-density uncertainties of the individual sub-bands. The broad overlap between the distributions indicates qualitative consistency, although the in-band and broad-band spectral indices probe different frequency ranges.

\begin{figure*}
    \centering
    \includegraphics[width=2\columnwidth]{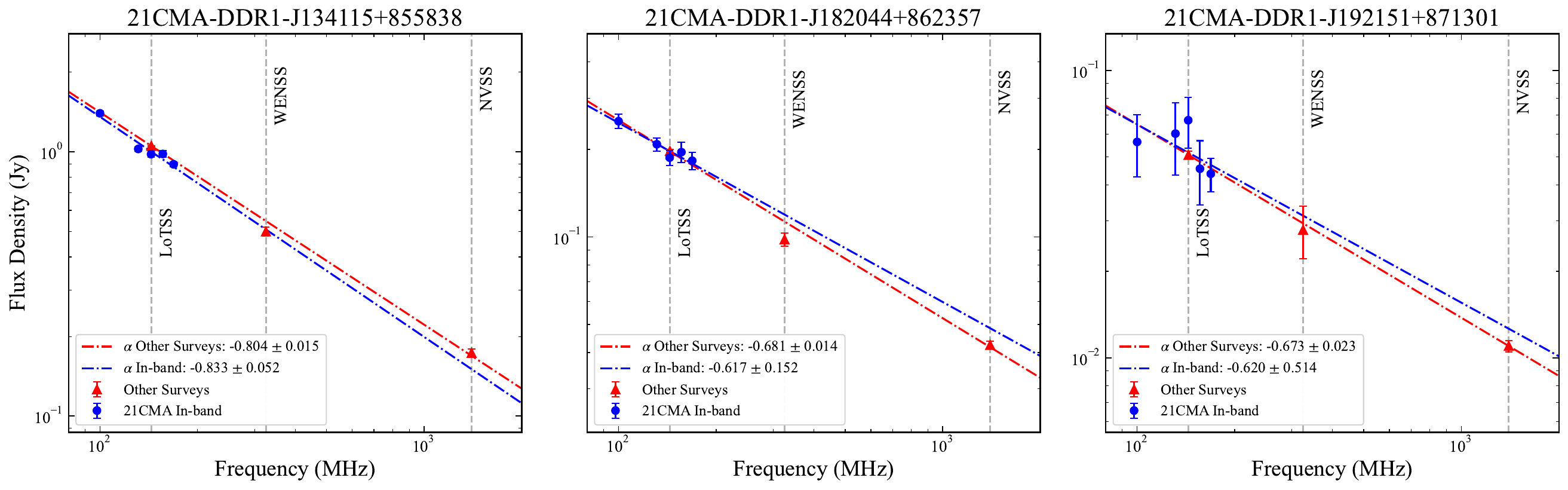}
    \caption{Spectral energy distributions (SEDs) of three representative 21CMA point sources spanning flux-density regimes of approximately 1.0\,Jy, 0.2\,Jy, and 0.05\,Jy at 150\,MHz, respectively. Blue dots represent the five in-band flux-density measurements from the 21CMA, while red triangles denote the external catalogue measurements from LoTSS (144\,MHz), WENSS (325\,MHz), and NVSS (1.4\,GHz). The fitted spectral indices and their uncertainties for the different data combinations are listed in the legends.}
    \label{fig:sed_comparison}
\end{figure*}

To illustrate this comparison on a source-by-source basis, Figure \ref{fig:sed_comparison} shows the SEDs of three representative point sources spanning different flux-density regimes ($\sim 1.0\,\mathrm{Jy}$, $\sim 0.2\,\mathrm{Jy}$, and $\sim 0.05\,\mathrm{Jy}$ at 150\,MHz, respectively). For the brighter sources, the five 21CMA sub-band measurements align closely with the broad-band power-law trend defined by the external surveys. As the source flux approaches the detection limit of the array, the uncertainty in the in-band fit, $\sigma_\alpha$, increases, mainly because the measurements become more strongly affected by thermal noise at lower signal-to-noise ratios. Even so, at flux densities of only a few tens of mJy, the in-band spectra derived from the 21CMA data remain broadly consistent with the external broad-band SEDs. This provides a qualitative consistency check on the 21CMA flux-density measurements across the 75--175\,MHz band.

\subsection{Noise Properties}
\label{subsec:noise_properties}

The sensitivity of a radio interferometer is fundamentally limited by thermal noise. Using the standard interferometric radiometer equation and the instrumental parameters adopted in earlier 21CMA studies \citep[e.g.,][]{Huang2016, Zheng2016}, the expected thermal-noise limit, $S_{\mathrm{thermal}}$, can be written as
\begin{equation}
    S_{\mathrm{thermal}} = \frac{\sqrt{2}\, k_{B} T_{\mathrm{sys}}}{A_{\mathrm{eff}} \eta \sqrt{\Delta\nu \Delta t}} \, ,
    \label{eq:thermal_noise}
\end{equation}
The present observations contain only the XX visibility product; Equation~\ref{eq:thermal_noise} therefore corresponds to a radio interferometer measuring a single linear polarisation. Here $k_{B}$ is the Boltzmann constant. The system temperature, $T_{\mathrm{sys}}$, includes both the receiver contribution ($T_{\mathrm{rcv}} \approx 50\,\mathrm{K}$) and the frequency-dependent sky temperature. At low frequencies, the latter is dominated by Galactic synchrotron emission and may be approximated as $T_{\mathrm{sky}}(\nu) \approx 60 (\nu / 300\,\mathrm{MHz})^{-2.55}\,\mathrm{K}$. Here, $\Delta\nu$ and $\Delta t$ denote the effective bandwidth and total integration time, respectively, with $\Delta t = 24$\,h for the present data set. The total effective collecting area is taken as $A_{\mathrm{eff}} = \sqrt{N_{\mathrm{bl}}}\, a_{\mathrm{eff}}$, where $a_{\mathrm{eff}} \approx 216\,\mathrm{m}^2$ is the effective area of a single station. After baseline selection and flagging, we retained $N_{\mathrm{bl}} = 728$ valid cross-correlation baselines. The system efficiency factor was set to $\eta = 0.5$ to account for correlator and signal-combination losses.

To measure the background noise in the restored images, we imposed a strict spatial restriction in the analysis. In wide-field radio interferometry, attenuation towards the edge of the primary beam leads to a substantial amplification of the background noise after primary-beam correction. To evaluate the core instrumental performance and the local rms noise level while minimising these edge effects, we therefore restricted the statistics to the central region within a radius of $3^\circ$ (i.e. $\mathrm{Dec} > 87^\circ$). Within this region, we used \textsc{PyBDSF} to estimate the local rms noise, enabling the adaptive rms box and spatially varying PSF model to capture non-Gaussian fluctuations and residual artefacts associated with synthesised-beam sidelobes around bright sources. Here we characterise the image noise using the median \textsc{PyBDSF} local rms within the central $3^\circ$ region ($\mathrm{Dec}>87^\circ$), while the residual-image rms values in Table~\ref{tab:final_image_properties} were derived separately from sigma-clipped statistics over $\mathrm{Dec}>85^\circ$.

\begin{figure*}
    \centering
    \includegraphics[width=2\columnwidth]{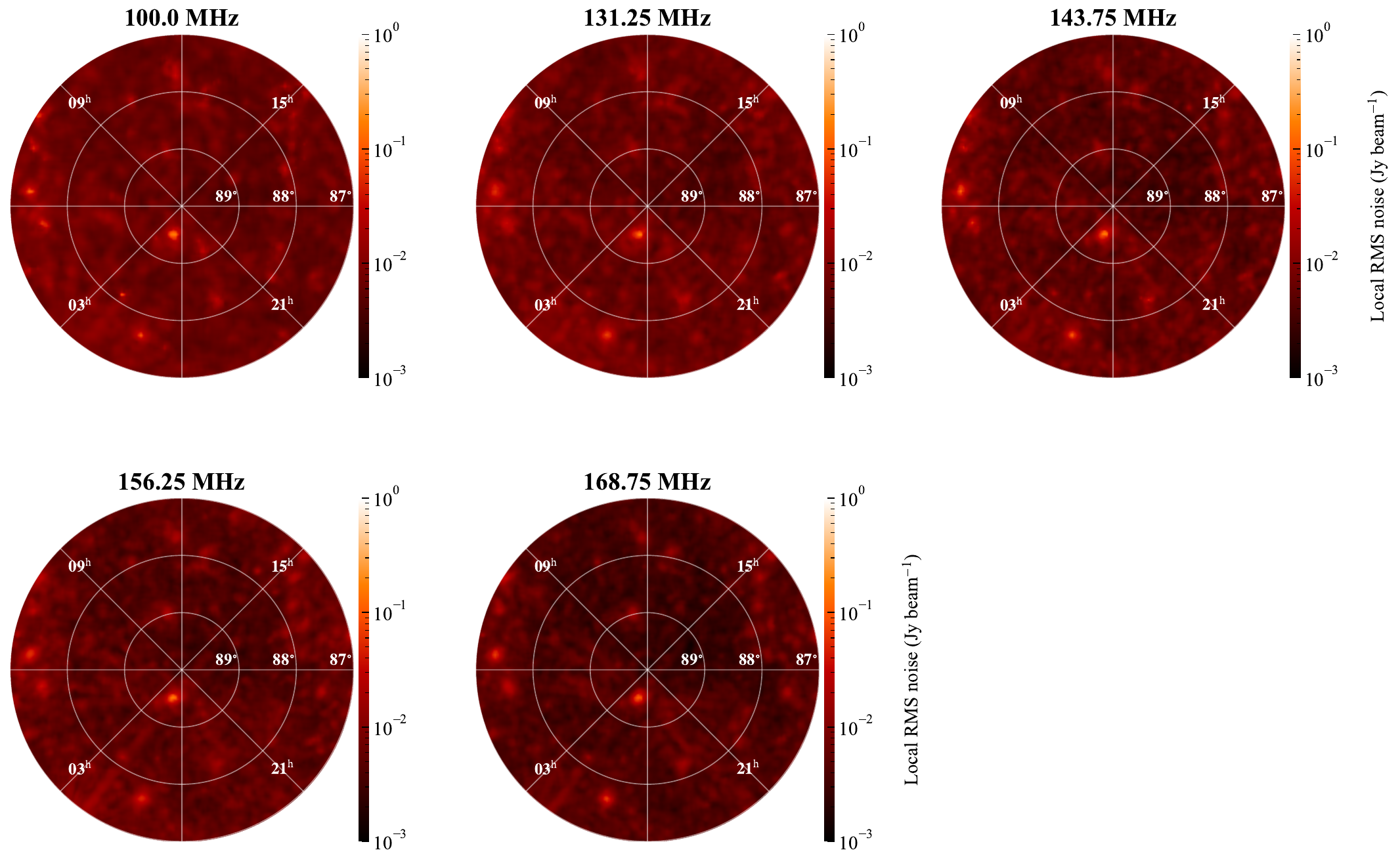}
    \caption{Spatial distribution of the local rms noise in the five 21CMA sub-bands within the central $3^\circ$ field ($\mathrm{Dec} > 87^\circ$). All panels use the same logarithmic colour scale.}
    \label{fig:rms_maps}
\end{figure*}

\begin{figure}
    \centering
    \includegraphics[width=\columnwidth]{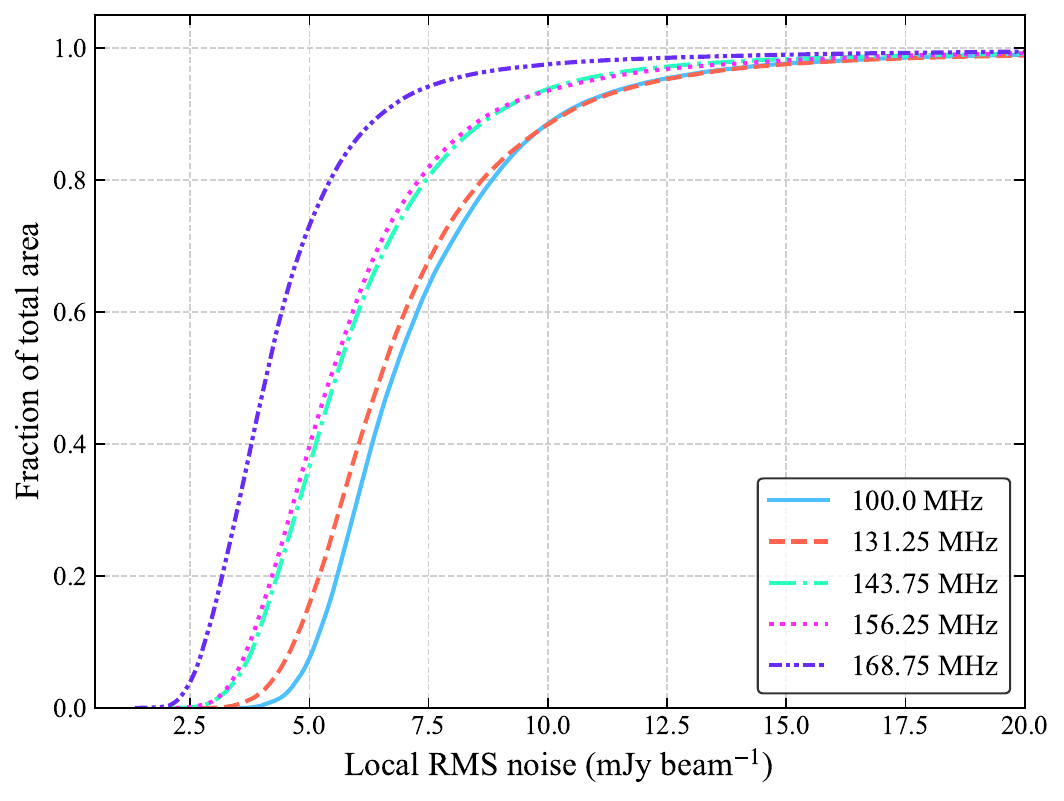}
    \caption{Cumulative area fraction as a function of the local rms noise in the central $3^\circ$ field ($\mathrm{Dec} > 87^\circ$) for the five frequency bands. The horizontal axis shows the local rms noise threshold, while the vertical axis gives the fraction of the total area with local noise below that threshold.}
    \label{fig:rms_cumulative}
\end{figure}

Figure \ref{fig:rms_maps} shows the spatial distribution of the local rms noise within the central $3^\circ$ field for the five sub-bands, while Figure \ref{fig:rms_cumulative} gives the corresponding cumulative area-fraction curves. As summarised in Table \ref{tab:noise_properties}, both the theoretical thermal noise and the measured median rms show a general decline towards higher frequencies, mainly driven by the rapid decrease in the low-frequency sky background temperature. The 100\,MHz band is particularly notable. Although it has an exceptionally large system temperature ($T_{\mathrm{sys}} \approx 1038\,\mathrm{K}$), its unusually wide synthesised bandwidth ($\Delta\nu = 50\,\mathrm{MHz}$) keeps the theoretical thermal noise relatively low at $0.335\,\mathrm{mJy\,beam^{-1}}$, slightly below that of the 131.25\,MHz band. Its measured median rms, however, reaches the highest value among all bands, at $6.74\,\mathrm{mJy\,beam^{-1}}$.

We estimated the classical-confusion noise using the restoring-beam dimensions of the final images. For each elliptical Gaussian restoring beam, we defined an equivalent circular FWHM as $\theta_{\mathrm{eff}}=(\theta_{\mathrm{maj}}\theta_{\mathrm{min}})^{1/2}$. The resulting values of $\theta_{\mathrm{eff}}$ are 2.168, 1.760, 1.616, 1.515, and 1.405\,arcmin at 100, 131.25, 143.75, 156.25, and 168.75\,MHz, respectively.

For a power-law differential source count $dN/dS\propto S^{-\gamma}$, the classical-confusion noise scales as $\sigma_{\mathrm{c}}\propto\theta^{2/(\gamma-1)}$ for a Gaussian beam \citep{Condon2012}. The 154-MHz source-count fit of \citet{Franzen2019} (their equation 11) gives a local slope of approximately $\gamma\simeq1.6$--$1.8$ over the flux-density range $S\simeq10$--$100\,\mathrm{mJy}$. The slope relevant to classical confusion is evaluated near $S_0=q\sigma_{\mathrm{c}}$, where $q\simeq5$ is commonly adopted for reliable source detection \citep{Condon2012}. For the $2\arcmin$ beam of \citet{Bernardi2009}, $\sigma_{\mathrm{c}}\simeq3\,\mathrm{mJy\,beam^{-1}}$ corresponds to $S_0\simeq15\,\mathrm{mJy}$, within this flux-density range. We therefore adopt $\gamma=1.6$ as a representative approximation, as also used in the completeness simulations, giving $\sigma_{\mathrm{c}}\propto\theta^{10/3}$ for beam sizes of order $2\arcmin$.

\citet{Bernardi2009} estimated this confusion level from WSRT observations at 150\,MHz by subtracting sources down to approximately $15\,\mathrm{mJy}$ and measuring an rms of $\sim3\,\mathrm{mJy\,beam^{-1}}$ in regions largely free of instrumental and ionospheric residuals. They also found this value consistent with an extrapolation of the 1.4-GHz confusion noise measured in the Hubble Deep Field, assuming an average spectral index of $\alpha=-0.8$. Using their reference value for a $2\arcmin$ beam and adopting our representative spectral index $\alpha=-0.712$, we estimated

\begin{equation}
\sigma_{\mathrm{c}}(\nu)
=
3
\left[
\frac{\theta_{\mathrm{eff}}(\nu)}
{2\arcmin}
\right]^{10/3}
\left(
\frac{\nu}{150\,\mathrm{MHz}}
\right)^{-0.712}
\mathrm{mJy\,beam^{-1}}.
\label{eq:classical_confusion}
\end{equation}

The resulting classical-confusion estimates are 5.24, 2.15, 1.52, 1.16, and $0.85\,\mathrm{mJy\,beam^{-1}}$, respectively.
Since the local source-count slope varies with flux density, these values should be regarded as approximate estimates rather than precise confusion limits, particularly for the smaller restoring beams.

\begin{table}
    \centering
    \caption{Noise properties of the five 21CMA frequency bands. The thermal-noise values correspond to the single XX linear-polarisation product. The classical-confusion values are estimated from the measured restoring beams using Equation~\ref{eq:classical_confusion}. The measured rms values are the median \textsc{PyBDSF} local rms within the central $3^\circ$ field ($\mathrm{Dec}>87^\circ$).}
    \label{tab:noise_properties}
    \footnotesize
    \setlength{\tabcolsep}{2.5pt}
    \begin{tabular}{cccccc}
        \hline
        \shortstack{Frequency\\(MHz)} & \shortstack{Bandwidth\\(MHz)} & \shortstack{$T_{\mathrm{sys}}$\\(K)} & \shortstack{Thermal\\noise} & \shortstack{Classical\\confusion} & \shortstack{Measured\\rms} \\
        & & & \multicolumn{3}{c}{($\mathrm{mJy\,beam^{-1}}$)} \\
        \hline
        100.00 & 50.0 & 1038.1 & 0.335 & 5.24 & 6.74 \\
        131.25 & 12.5 & 543.9 & 0.351 & 2.15 & 6.48 \\
        143.75 & 12.5 & 441.7 & 0.285 & 1.52 & 5.55 \\
        156.25 & 12.5 & 366.6 & 0.236 & 1.16 & 5.44 \\
        168.75 & 12.5 & 310.2 & 0.200 & 0.85 & 4.09 \\
        \hline
    \end{tabular}
\end{table}

Overall, the measured rms exceeds the theoretical thermal-noise limit at all five frequencies. At 100\,MHz, the estimated classical-confusion noise of $5.24\,\mathrm{mJy\,beam^{-1}}$ is comparable to the measured rms of $6.74\,\mathrm{mJy\,beam^{-1}}$, indicating that classical confusion is an important contribution, although the image cannot be regarded as limited by classical confusion alone. At 131.25--168.75\,MHz, the estimated classical-confusion levels of $0.85$--$2.15\,\mathrm{mJy\,beam^{-1}}$ remain below the measured rms values of $4.09$--$6.48\,\mathrm{mJy\,beam^{-1}}$, suggesting that residual calibration errors, incomplete deconvolution, and sidelobe-related artefacts may also contribute to the image noise. The particularly high rms in the 100\,MHz band is also physically plausible. In addition to the general challenges of low-frequency calibration, this band is more strongly affected by ionospheric phase distortions and by residual contamination from RFI near the FM band (88-108\,MHz). These effects may also help explain why the 100\,MHz band, despite not having the highest theoretical thermal-noise limit, exhibits the largest measured rms.

A comparison with earlier 21CMA processing results nevertheless highlights the substantial progress made in reducing residual calibration errors and imaging artefacts at the low-frequency end. At the higher frequencies around $\sim 150\,\mathrm{MHz}$, the measured noise in the present catalogue is comparable to the levels reported by \citet{Zheng2016} and \citet{Zhao2022}, all being of order $4\text{--}6\,\mathrm{mJy\,beam^{-1}}$. The classical-confusion estimates in Table~\ref{tab:noise_properties} nevertheless remain below the measured rms values at these frequencies, suggesting that additional calibration and imaging residuals likely contribute to the measured rms. By contrast, at the lowest frequencies around 100\,MHz, where ionospheric phase distortions are stronger and the observing band is more affected by residual RFI, the limitations of conventional calibration become much more apparent. \citet{Zheng2016} reported local rms values of $22.5\,\mathrm{mJy\,beam^{-1}}$ at 93.75\,MHz and $17.8\,\mathrm{mJy\,beam^{-1}}$ at 106.25\,MHz, while \citet{Zhao2022} reported a value of approximately $20\,\mathrm{mJy\,beam^{-1}}$ at 100\,MHz. In contrast, the present work yields a local rms of $6.74\,\mathrm{mJy\,beam^{-1}}$ in the 100\,MHz band. Although differences in bandwidth, integration time, and data selection should be kept in mind, this comparison still suggests a marked improvement in image quality at the low-frequency end.

Because classical confusion does not decrease with additional integration time, combining observations from additional days alone will not reduce this component of the image noise. Reducing the classical-confusion contribution would require a smaller synthesised beam provided by longer baselines, while further improvements in calibration, source modelling, and deconvolution will be required to reduce the remaining excess noise.

\subsection{Source Counts}
\label{subsec:source_counts}

\subsubsection{Completeness}
\label{subsec:completeness}

\begin{figure*}
    \centering
    \includegraphics[width=2\columnwidth]{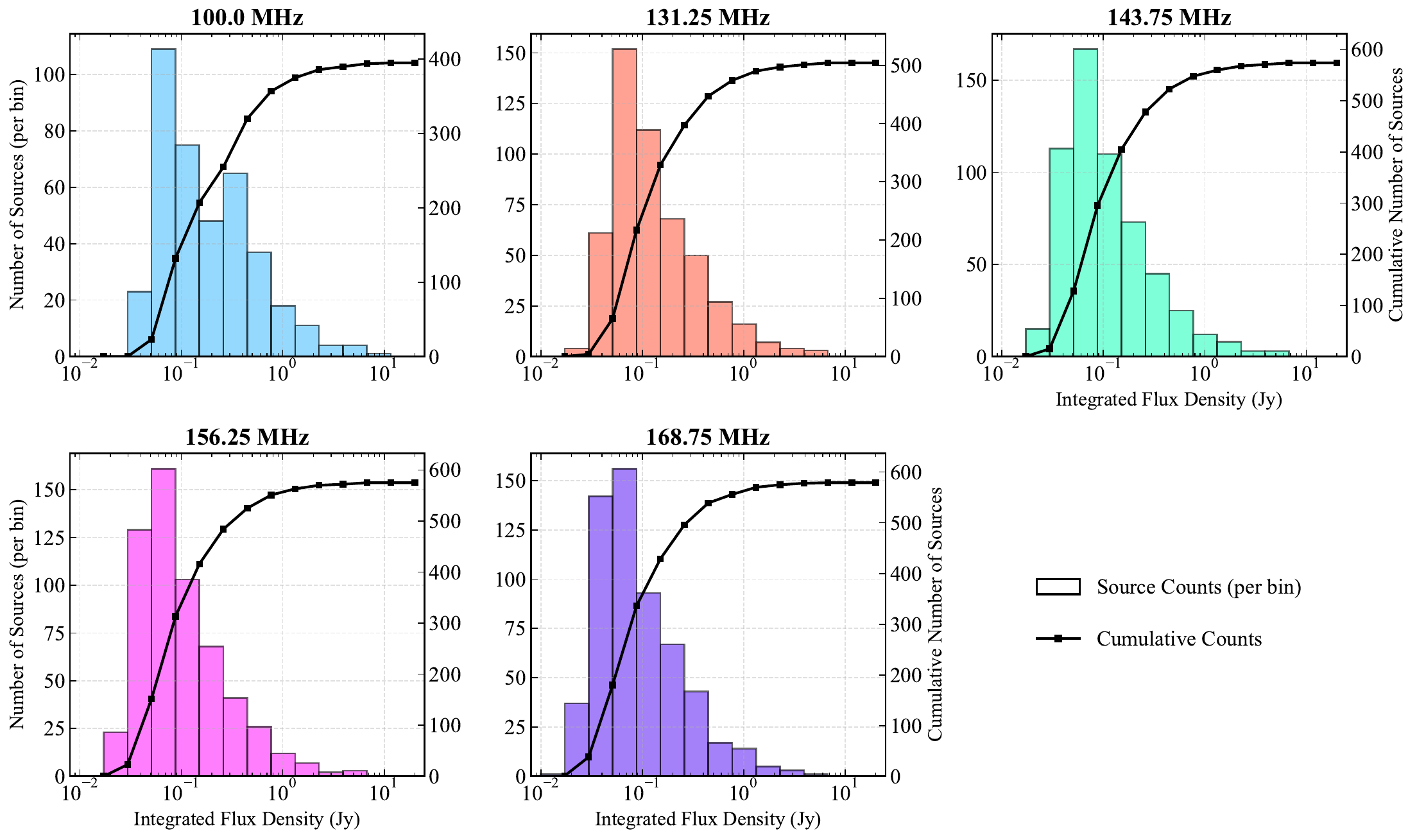}
    \caption{Distributions of the integrated flux densities of the 580 real point sources detected within the central $3^\circ$ core region ($\mathrm{Dec} > 87^\circ$). In each panel, the shaded histogram (left axis) shows the absolute number of sources in each logarithmic flux-density bin, while the solid line with square markers (right axis) gives the cumulative number of sources.}
    \label{fig:source_counts_hist}
\end{figure*}

Before deriving the final differential source counts, it is necessary to quantify the completeness of the catalogue. Because sensitivity attenuation towards the edge of the primary beam introduces non-Gaussian noise and systematic flux errors, we restricted the completeness analysis, and all subsequent source-count measurements, to the central $3^\circ$ region (i.e. $\mathrm{Dec} > 87^\circ$). After this spatial cut, the final catalogue contains 580 high-confidence point sources within the core region. Figure \ref{fig:source_counts_hist} shows the distributions of integrated flux density for these 580 sources in the five sub-bands. Most detected sources lie in the flux-density range $0.01$--$1\,\mathrm{Jy}$.

To assess the detection completeness in this flux regime, we employed a Monte Carlo injection analysis based on the residual maps \citep[e.g.,][]{LoTSS-2, LoTSS-5, GLEAM-1}. In these simulations, the \textsc{PyBDSF} residual maps obtained after real-source extraction were used as a realistic representation of the background noise. Artificial sources were injected according to a power-law flux-density distribution, $dN/dS \propto S^{-1.6}$ \citep[e.g.,][]{MSSS}, and were modelled as two-dimensional Gaussian components with shapes matched to the restoring beam of each sub-band. The injected flux densities at $150\,\mathrm{MHz}$ spanned $0.002$--$6\,\mathrm{Jy}$. The adopted slope is broadly representative of differential source counts near $150\,\mathrm{MHz}$ over $\sim10$--$100\,\mathrm{mJy}$ \citep{Franzen2019}, rather than the entire injected range. We carried out 500 independent realisations, injecting 300 mock sources at common random sky positions across the five sub-bands within the core region in each run, with flux densities scaled from $150\,\mathrm{MHz}$ assuming a fixed spectral index of $\alpha=-0.825$. This number was chosen to provide sufficient Monte Carlo statistics while avoiding any significant modification of the source crowding and background properties of the original images. Blind source extraction was then performed using the same \textsc{PyBDSF} detection parameter settings as for the real data. For each sub-band, completeness was calculated as the fraction of injected sources recovered in that sub-band and detected in at least three of the five sub-bands, following the catalogue selection criterion described in Section~\ref{subsec:spectral_fitting}.

\begin{figure*}
    \centering
    \includegraphics[width=2\columnwidth]{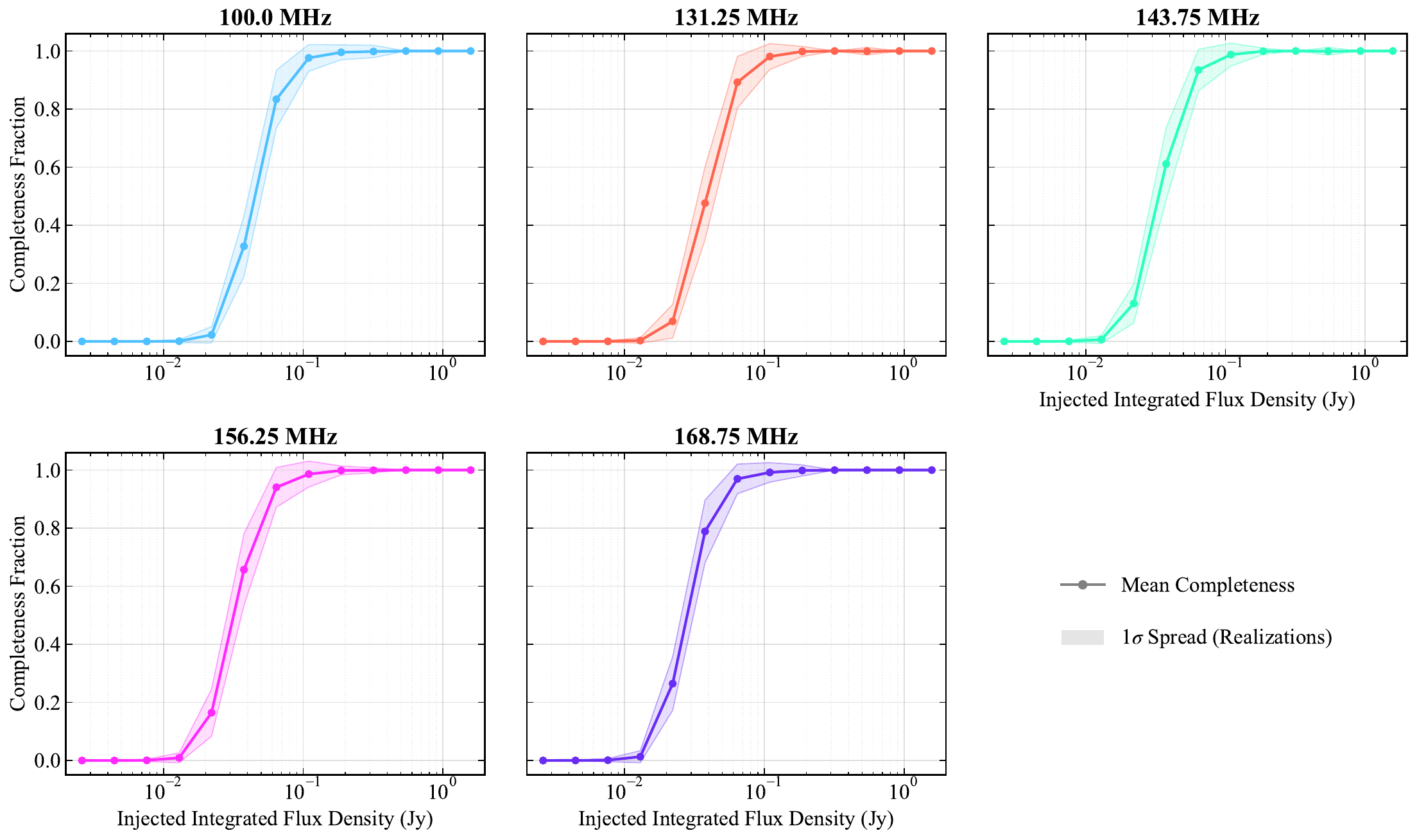}
    \caption{Detection completeness as a function of the injected integrated flux density for the five 21CMA sub-bands, derived from joint Monte Carlo injection--recovery simulations incorporating the three-sub-band detection criterion. The simulations were restricted to the central $3^\circ$ core region ($\mathrm{Dec} > 87^\circ$). The solid lines with circles show the mean completeness from 500 independent realisations, while the shaded regions indicate the corresponding $1\sigma$ standard deviations.}
    \label{fig:completeness}
\end{figure*}

Figure \ref{fig:completeness} shows the detection completeness derived from the Monte Carlo simulations as a function of the injected integrated flux density for the five 21CMA sub-bands within the core region. All completeness curves display the expected smooth sigmoidal shape. With increasing frequency, the curves shift systematically towards lower flux densities, indicating that the higher frequency bands reach full completeness at lower flux densities than the lower frequency bands. This frequency dependence is consistent with the decrease in local background noise discussed in the previous section. Together with Figure~\ref{fig:source_counts_hist}, these results indicate high completeness for the brighter detected sources, with increasing incompleteness towards the faint end of the catalogue.

\subsubsection{Differential Source Counts}
\label{subsec:differential_counts}

\begin{figure}
    \centering
    \includegraphics[width=\columnwidth]{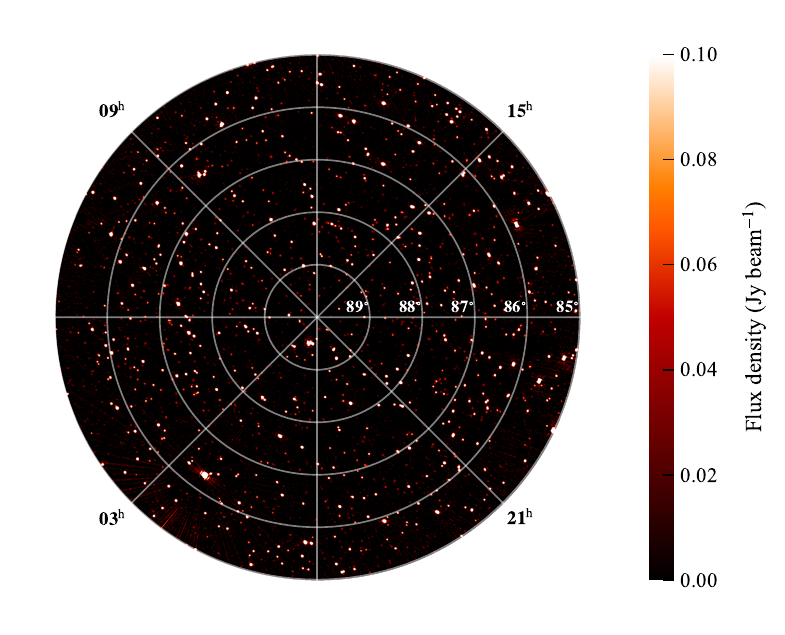}
    \caption{Primary-beam-corrected 150\,MHz wide-band MFS image generated from the four DD-calibrated sub-bands covering 125--175\,MHz. This image was used for source extraction in the differential source-count analysis.}
    \label{fig:150MHz_MFS}
\end{figure}

To validate the flux-density calibration of the 21CMA images from a statistical perspective, and to examine the population distribution of low-frequency radio sources, we calculated the Euclidean-normalised differential source counts. The analysis was restricted to the central $3^\circ$ core region ($\mathrm{Dec} > 87^\circ$) to minimise the impact of primary-beam attenuation and spatially varying noise. For this analysis, we generated a dedicated 50-MHz-wide MFS image centred at 150\,MHz from the four DD-calibrated sub-bands covering 125--175\,MHz. Source extraction, astrometric correction, and flux-density rescaling were performed using the same procedures adopted for the main catalogue. The independently extracted 150-MHz source list was then cross-matched with the final 21CMA-DDR1 catalogue, and the 580 matched sources within the core region formed the statistical sample. The 150-MHz flux densities used here were therefore measured from this dedicated MFS image rather than taken from the multi-band catalogue. The resulting image is shown in Figure~\ref{fig:150MHz_MFS}.

The integrated flux densities were binned logarithmically over the range $0.017$--$5.95\,\mathrm{Jy}$. The Euclidean-normalised differential source count is defined as
\begin{equation}
    \Delta N_{\mathrm{norm}} = \frac{N}{\Omega \Delta S} S_c^{2.5}\, ,
\end{equation}
where $N$ is the number of sources in a given flux-density bin, $\Delta S$ is the bin width, $S_c$ is the geometric centre of the bin, and $\Omega$ is the solid angle of the core statistical region ($\Omega \approx 0.0086\,\mathrm{sr}$). To present the actual detection limits and the baseline performance of the 21CMA wideband images, we did not apply any explicit corrections for completeness or Eddington bias. The uncertainty in each bin is assumed to be dominated by Poisson statistics, such that $\sigma_{\mathrm{norm}} = \Delta N_{\mathrm{norm}} / \sqrt{N}$. The detailed parameters and the resulting source counts for each bin are listed in Appendix \ref{appendix:source_counts_table}.

\begin{figure*}
    \centering
    \includegraphics[width=2\columnwidth]{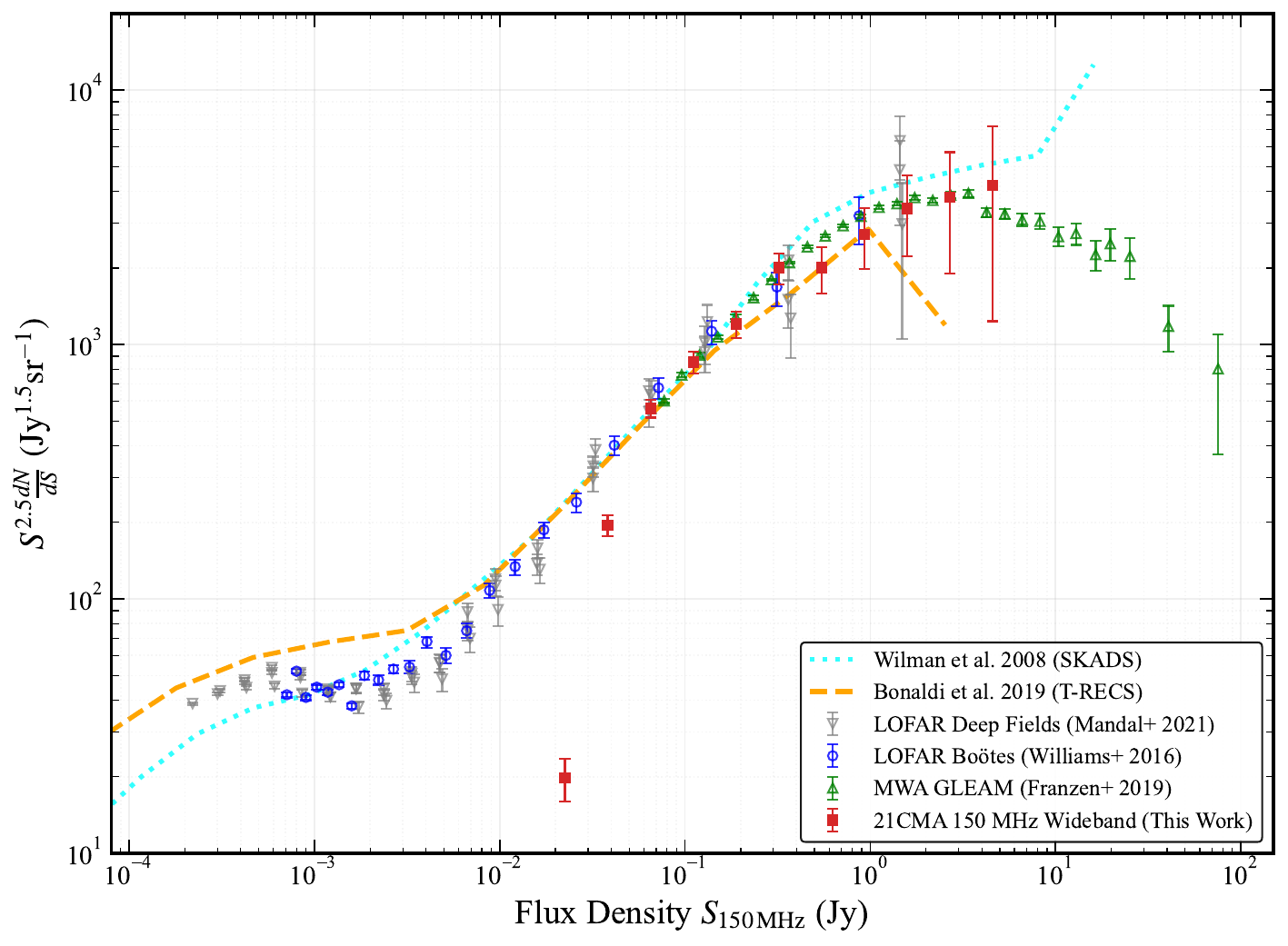}
    \caption{Euclidean-normalised differential source counts at $150\,\mathrm{MHz}$. The red squares represent the source counts derived from the dedicated 21CMA 150\,MHz wide-band MFS image within the central $3^\circ$ core region ($\mathrm{Dec} > 87^\circ$). For comparison, representative literature measurements are also shown, namely the LOFAR Deep Fields (grey downward triangles; \citealt{e-curve-lotss}), LOFAR Bo\"{o}tes (blue circles; \citealt{e-curve-bootes}), and MWA GLEAM (green upward triangles; \citealt{Franzen2019}). The cyan dotted and orange dashed curves denote the predictions of the SKADS \citep{SKADS} and T-RECS \citep{T-RECS} simulation models, respectively.}
    \label{fig:source_counts}
\end{figure*}

To test the 21CMA source counts against a broader observational context, we compared our results at $150\,\mathrm{MHz}$ with published observational measurements and simulation models. In Figure \ref{fig:source_counts}, we include three representative external surveys: the MWA GLEAM survey at $154\,\mathrm{MHz}$ for the bright-flux regime \citep{Franzen2019}, the LOFAR Bo\"{o}tes observations at $150\,\mathrm{MHz}$ for the intermediate regime \citep{e-curve-bootes}, and the LOFAR Deep Fields at $150\,\mathrm{MHz}$ for the faintest regime \citep{e-curve-lotss}. We also include two widely used semi-empirical radio-sky simulation models, SKADS \citep{SKADS} and T-RECS \citep{T-RECS}.

As shown in Figure \ref{fig:source_counts}, the overall shape of the 21CMA wideband differential source counts is consistent with both the external observational results and the simulation models, including the broad Euclidean-bump feature commonly associated with the radio-loud AGN population. In terms of absolute normalisation, however, the 21CMA source counts show a slight systematic offset towards lower values. This modest global offset may be related in part to our catalogue-construction strategy. As described in Section \ref{subsec:strategy}, we applied strict morphological filtering to exclude complex extended sources in order to construct a high-purity point-source catalogue. By contrast, the external surveys and simulation models include radio sources of all morphological types. A small systematic deficit relative to those total-population counts is therefore expected. Residual flux-density-scale differences described in Section~\ref{sec:flux_val} may also contribute to the normalisation offset.

At integrated flux densities $S \lesssim 0.04\,\mathrm{Jy}$, the 21CMA source counts show a clear downturn, and the deficit relative to the external measurements becomes more pronounced. This behaviour is consistent with the completeness curves shown in Figure \ref{fig:completeness}, which decline rapidly at comparable flux densities. This faint-end suppression does not imply a physical decline in the true source density, but instead reflects the detection limits of the present data and sample selection. First, limited surface-brightness sensitivity, together with classical confusion, prevents a fraction of faint compact sources from reaching the $5\sigma$ peak detection threshold. Secondly, the statistical sample is required to be associated with the final 21CMA-DDR1 catalogue, which in turn requires valid detections in at least three independent sub-bands. Although this multi-band criterion improves the reliability of the catalogue, it inevitably excludes some marginal sources close to the detection limit. The observed downturn therefore provides an empirical indication of the effective completeness boundary of the present 24-h data in the core region, and offers a useful reference for deeper analyses based on combined multi-day 21CMA data.

\section{Conclusions and Future Work}
\label{sec:conclusion}

In this paper, we presented the 21CMA-DDR1 catalogue, a low-frequency radio point-source catalogue of the NCP region based on a continuous 24-h 21CMA observation. Using the DDC and wide-field imaging pipeline \texttt{21CMAcali}, we recovered 1246 extragalactic radio point sources over the 75--175\,MHz band. The in-band spectral-index distribution has a median value of $\alpha=-0.825$, consistent with a synchrotron-dominated extragalactic source population. Cross-matching with external surveys, together with analyses of the positional offsets, noise properties, completeness, and differential source counts, showed that the resulting catalogue has high positional accuracy and a broadly consistent flux-density scale.

A remaining limitation of the present catalogue is the treatment of ultra-bright extended sources in the wide field. In particular, the dominant radio galaxy 3C\,61.1 still imposes a strong constraint on the achievable dynamic range. In this work, we improved its representation already at the stage of the initial calibration sky model by replacing the usual single-source approximation with two point-source components. This simplified treatment helped to reduce the associated grating-lobe contamination across the field and thereby improved the quality of the subsequent calibration and imaging, contributing to the recovery of a substantially larger source population. Nevertheless, this representation remains far from a full morphological description of the source, and does not yet reach the level of sophistication achieved in more advanced low-frequency imaging pipelines such as those developed for LOFAR. A natural next step will therefore be to introduce more flexible morphological models, for example based on \textsc{Shapelet} basis functions \citep[e.g.,][]{shapelet-1,shapelet-2}, in order to better represent the extended surface-brightness structure of such sources. This should further reduce residual artefacts and improve the accuracy of the direction-dependent calibration.

The validation of the automated DDC pipeline on a single-day data set provides a practical basis for future processing of deeper 21CMA observations. A natural next step is to apply \texttt{21CMAcali} to combined multi-day data in order to improve the depth and fidelity of the final images. With longer effective integration time, the thermal-noise contribution will be reduced, while improved modelling of bright extended sources, as discussed above, should further suppress residual artefacts. This combination should improve the recovery of fainter radio sources, while further gains in sensitivity will depend on reducing classical confusion and residual calibration and imaging errors. In particular, the use of longer baselines would provide a smaller synthesised beam and help extend the catalogue to lower flux densities. Another possible improvement is to refine the temporal sampling of the DD calibration solutions. The present 5-min complex-gain solution interval may average over faster ionospheric phase variations \citep{Vedantham2015}. Future processing could combine shorter ($\sim10$--$60$\,s), broader-band phase-only or total electron content (TEC)-constrained solutions with slower, narrow-band complex-gain calibration.

From a broader scientific perspective, this work marks an important step towards the foreground characterisation required for future 21-cm EoR power-spectrum studies with the 21CMA. Although the present pipeline and catalogue still fall short of the precision ultimately needed for a robust cosmological detection, they establish a practical path for improving low-frequency calibration, imaging, and sky modelling, thereby helping to advance future 21CMA power-spectrum analyses and to accumulate useful experience for SKA-era low-frequency cosmology.

\section*{Acknowledgements}

We thank the anonymous referee for constructive comments
and suggestions that helped improve the manuscript. We thank Cyril Tasse, Benjamin Hugo and Oleg Smirnov for helpful discussions on the practical use of \textsc{DDFacet} and \textsc{killMS}, and on direction-dependent calibration strategies. We also thank Sarvesh Mangla for helpful discussions on point-source detection, Zhaoting Chen for constructive suggestions on the data analysis, Aishrila Mazumder for helpful suggestions on the validation analyses and manuscript revisions, and Zifan Zhang for suggestions on the colour schemes used in the statistical figures. The main data processing for this work was carried out on the SKATS computing server at the Shanghai Astronomical Observatory.

This work was supported by the National SKA Program of China (No. 2025SKA0160100) and the National Natural Science Foundation of China (NSFC; grant Nos. 12573112, 12573097, and 12673113).

This research made use of the following radio astronomy software packages: \textsc{CASA} \citep{CASA}, \textsc{killMS} \citep{2015MNRAS.449.2668S}, \textsc{DDFacet} \citep{DDFacet}, and \textsc{PyBDSF} \citep{PyBDSF}. It also made use of the Python scientific computing packages \textsc{Astropy} \citep{astropy-1, astropy-2, astropy-3} and \textsc{NumPy} \citep{numpy}.


\section*{Data Availability}

The full machine-readable 21CMA-DDR1 catalogue is provided as online supplementary material accompanying this article. The calibrated images and associated data products underlying this work are available from the corresponding author upon reasonable request.




\bibliographystyle{mnras}
\bibliography{example} 




\appendix

\section{Catalogue Column Descriptions}
\label{appendix:catalogue_columns}

The full 21CMA-DDR1 catalogue is available in machine-readable form as online supplementary material. It contains 1246 high-confidence extragalactic radio point sources and 16 data columns. The column definitions and units are as follows:

\begin{itemize}
    \item \textbf{Column 1: Source\_ID} -- Unique source identifier following the standard IAU convention, \texttt{21CMA-DDR1-JHHMMSS+DDMMSS}, derived from the J2000 coordinates.
    \item \textbf{Columns 2--3: R. A. (J2000), Dec. (J2000)} -- Right Ascension and Declination of the source in the J2000 frame, given in hours, minutes, and seconds (h:m:s) and degrees, arcminutes, and arcseconds (d:m:s), respectively. These positions have been corrected for systematic offsets using the NVSS astrometric reference frame.
    \item \textbf{Column 4: Source\_Type} -- Morphological classification of the radio source. Because this work is restricted to unresolved emission (see Section \ref{subsec:strategy}), all entries are labelled `P' (point source).
    \item \textbf{Columns 5--14: F$\nu$, e\_F$\nu$} -- Integrated flux densities and their corresponding $1\sigma$ uncertainties for the five 21CMA sub-bands ($\nu \in \{100, 131.25, 143.75, 156.25, 168.75\}\,\mathrm{MHz}$), in units of Jy. Entries are left blank where no valid local detection is available.
    \item \textbf{Columns 15--16: Spectral\_Index, e\_Spectral\_Index} -- Best-fitting in-band spectral index $\alpha$ and its uncertainty, derived from logarithmic linear regression ($\ln S \propto \alpha \ln \nu$).
\end{itemize}

\section{Differential Source Counts Table}
\label{appendix:source_counts_table}

Table \ref{tab:source_counts} lists the Euclidean-normalised differential source counts derived from the dedicated 21CMA 150\,MHz wide-band MFS image described in Section~\ref{subsec:differential_counts}. To ensure statistical reliability and minimise systematic effects related to primary-beam attenuation, the analysis is restricted to the central $3^\circ$ core region ($\mathrm{Dec} > 87^\circ$).

\begin{table}
    \centering
    \caption{Euclidean-normalised differential source counts derived from the dedicated 21CMA 150\,MHz wide-band MFS image within the core region ($\mathrm{Dec} > 87^\circ$). The quoted uncertainties, $\sigma_{\mathrm{norm}}$, are based on Poisson statistics.}
    \label{tab:source_counts}
    \begin{tabular}{cccccc}
        \hline
        \shortstack{$S_{\rm start}$\\(Jy)} & \shortstack{$S_{\rm end}$\\(Jy)} & \shortstack{$S_c$\\(Jy)} & $N_{\rm sources}$ & \shortstack{$\Delta N_{\rm norm}$\\($\mathrm{Jy}^{1.5}\,\mathrm{sr}^{-1}$)} & \shortstack{$\sigma_{\rm norm}$\\($\mathrm{Jy}^{1.5}\,\mathrm{sr}^{-1}$)} \\
        \hline
        0.017 & 0.029 & 0.023 & 27  & 19.76   & 3.80 \\
        0.029 & 0.050 & 0.038 & 120 & 194.76  & 17.78 \\
        0.050 & 0.085 & 0.065 & 156 & 561.57  & 44.96 \\
        0.085 & 0.145 & 0.111 & 107 & 854.33  & 82.59 \\
        0.145 & 0.246 & 0.189 & 68  & 1204.24 & 146.04 \\
        0.246 & 0.418 & 0.321 & 51  & 2003.27 & 280.51 \\
        0.418 & 0.712 & 0.546 & 23  & 2003.83 & 417.83 \\
        0.712 & 1.210 & 0.928 & 14  & 2705.35 & 723.03 \\
        1.210 & 2.059 & 1.578 & 8   & 3428.85 & 1212.28 \\
        2.059 & 3.501 & 2.685 & 4   & 3802.61 & 1901.30 \\
        3.501 & 5.954 & 4.566 & 2   & 4217.11 & 2981.95 \\
        \hline
    \end{tabular}
\end{table}




\bsp	
\label{lastpage}
\end{document}